\documentclass[twocolumn]{aastex63}
\turnoffedit
\usepackage{amsmath}
\usepackage{chngcntr}
\usepackage{stix}
\usepackage{enumitem}
\usepackage{xspace}
\newcommand{\halpha}{H\ensuremath{\alpha}\xspace}
\newcommand{\angstrom}{\text{\normalfont\AA}}

\newcommand{\Lacc}{\ensuremath{L_\mathrm{acc}}\xspace}
\newcommand{\Lline}{\ensuremath{L_\mathrm{line}}\xspace}
\newcommand{\Lhalpha}{\ensuremath{L_\mathrm{H\alpha}}\xspace}
\newcommand{\Lcont}{\ensuremath{L_\mathrm{cont}}\xspace}
\newcommand{\Lsol}{\ensuremath{L_{\odot}}\xspace}
\graphicspath{{./}{figures/}{Figures/}}

\shorttitle{PDS 70 \lowercase{b} in UV and \halpha}
\shortauthors{Zhou et al.}

\begin{document}
\title{Hubble Space Telescope UV and H$\alpha$ Measurements of the Accretion Excess Emission\\ from the Young Giant Planet PDS~70~b}
\correspondingauthor{Yifan Zhou}
\email{yifan.zhou@utexas.edu}
\author[0000-0003-2969-6040]{Yifan Zhou}
\altaffiliation{Harlan J. Smith McDonald Observatory Fellow}
\affiliation{Department of Astronomy/McDonald Observatory, The University of Texas,  2515 Speedway, Stop C1400 Austin, TX 78712, USA}
\author[0000-0003-2649-2288]{Brendan P. Bowler}
\affiliation{Department of Astronomy/McDonald Observatory, The University of Texas,  2515 Speedway, Stop C1400 Austin, TX 78712, USA}
\author[0000-0002-4309-6343]{Kevin R. Wagner}
\altaffiliation{NASA Hubble Fellow}
\affiliation{Department of Astronomy/Steward Observatory, The University of Arizona, 933 N Cherry Ave Tucson, AZ 85721, USA}
\author[0000-0002-4511-5966]{Glenn Schneider}
\affiliation{Department of Astronomy/Steward Observatory, The University of Arizona, 933 N Cherry Ave Tucson, AZ 85721, USA}
\author[0000-0003-3714-5855]{D\'aniel Apai}
\affiliation{Department of Astronomy/Steward Observatory, The University of Arizona, 933 N Cherry Ave Tucson, AZ 85721, USA}
\affiliation{Department of Planetary Science/Lunar and Planetary Laboratory, The University of Arizona, 1640 E. University Blvd., Tucson, AZ 85721, USA}
\author[0000-0001-9811-568X]{Adam L. Kraus}
\affiliation{Department of Astronomy/McDonald Observatory, The University of Texas,  2515 Speedway, Stop C1400 Austin, TX 78712, USA}
\author[0000-0002-2167-8246]{Laird M. Close}
\affiliation{Department of Astronomy/Steward Observatory, The University of Arizona, 933 N Cherry Ave Tucson, AZ 85721, USA}
\author[0000-0002-7154-6065]{Gregory J. Herczeg}
\affiliation{Kavli Institute for Astronomy and Astrophysics, Peking University, Yi He Yuan Lu 5, Haidian Qu, Beijing 100871, China}
\author[0000-0001-8060-1321]{Min Fang}
\affiliation{Purple Mountain Observatory, Chinese Academy of Sciences, Nanjing 210023, China}
\newcommand{\pds}{PDS 70\xspace}
\newcommand{\pdsb}{PDS 70 b\xspace}
\newcommand{\pdsbc}{PDS 70 b and c\xspace}
\newcommand{\unfinished}[1]{{\color{red}#1}}

\begin{abstract}
  Recent discoveries of young exoplanets within their natal disks offer exciting opportunities to study ongoing planet formation. In particular, a planet's mass accretion rate can be constrained by observing the accretion-induced excess emission. So far, planetary accretion is only probed by the \halpha line, \edit1{which is then converted to a total accretion luminosity using correlations derived for stars}. However, the  majority of the accretion luminosity is expected to emerge from hydrogen continuum emission, \edit1{and is best measured in the ultraviolet (UV). } In this paper, we present HST/WFC3/UVIS F336W (UV) and F656N (\halpha)  high-contrast imaging observations of PDS 70. Applying a suite of novel observational techniques, we detect the planet PDS 70 b with signal-to-noise ratios of 5.3 and 7.8 in the F336W and F656N bands, respectively. This is the first time that an exoplanet has been directly imaged in the UV. Our \edit1{observed \halpha flux of \pdsb{} } is higher  by $3.5\sigma$ than the \edit1{most recent published result. }  \edit1{However, the light curve retrieved from our observations does not support greater than 30\% variability in the planet's \halpha emission in six epochs over a five-month timescale. }  We estimate a mass accretion rate of $1.4\pm0.2\times10^{-8}M_{\mathrm{Jup}}\,\mathrm{yr}^{-1}$. \halpha accounts for 36\% of the total  accretion luminosity. Such a high proportion of energy released in line emission suggests efficient production of \halpha emission in planetary accretion, and motivates using the \halpha band for searches of accreting planets. These results demonstrate HST/WFC3/UVIS's excellent high-contrast imaging performance and highlight its potential for planet formation studies.
\end{abstract}

\section{Introduction}

Directly imaged protoplanets are excellent testbeds for planet formation theories. These planets reside in the gaps of circumstellar disks, supporting models in which the observed protoplanetary disk gaps \citep[e.g.,][]{Andrews2018} are carved by newly formed planets \citep[e.g.,][]{Dodson2011,Zhu2011,Zhang2018b,Bae2019}. Ongoing accretion has been detected in these planets through their \halpha  emission \citep{Sallum2015,Wagner2018,Haffert2019,Hashimoto2020}. The strength and velocity profile of the \halpha line can be used to indirectly constrain planetary mass accretion rates. These results enable quantitative investigations of accretion physics \citep[e.g.,][]{Aoyama2018,Thanathibodee2019,Szulagyi2020}, planet--disk interactions \citep[e.g.,][]{Dong2014,Bae2019}, and planetary luminosity evolution \citep[e.g.,][]{Marley2007,Marleau2017}.  \edit1{With its two directly imaged actively accreting planets, PDS 70 is an excellent target for planet formation studies.}

\pds is a ${\sim}{5}~\mathrm{Myr}$ old K7 T Tauri star in the Upper Sco association \citep{Pecaut2016}. The star is slowly accreting  \citep{Thanathibodee2020} and hosts a disk with complex structures including a giant inner cavity \citep{Hashimoto2012,Keppler2019}. Two planets, PDS~70 b and c, have been discovered within the disk cavity \citep{Keppler2018,Haffert2019} located at projected separations of 20 and 34 AU from the star, respectively.  \edit1{Using high-contrast imaging observations, \citet{Wagner2018}, \citet{Haffert2019}, and \citet{Hashimoto2020} probed accretion onto these planets using the \halpha{} line. In addition, \citet{Christiaens2019}, \citet{Stolker2020}, and \citet{Uyama2021} placed upper limits on infrared accretion tracers, such as the Br-$\gamma$ and Pa-$\beta$ lines. Based on these observations, accretion rates of \pds~b and c are estimated to be in the range of $1\times10^{-8}$ to $5\times10^{-7}$ $M_{\mathrm{Jup}}\,\mathrm{yr^{-1}}$\citep[e.g.,][]{Wagner2018, Haffert2019,Aoyama2019,Thanathibodee2019, Hashimoto2020,Aoyama2020}.}


\edit1{The accretion process is usually modeled as material falling onto the star or planet at nearly freefall velocity \citep[e.g.,][]{Calvet1998, Hartmann2016, Aoyama2018, Aoyama2020}. As the accretion flow hits
the stellar/planetary surface, its kinetic energy is converted into thermal energy, forming a hot shock front. In the magnetospheric accretion model of T Tauri stars, the shock front heats up its surrounding gas, which emits the majority of the accretion luminosity in the form of hydrogen continuum emission \citep[e.g.,][]{Calvet1998,Hartmann2016}. The accretion columns, which are less dense than the shock front and optically thin in the hydrogen lines, produce broad emission lines  \citep[e.g.,][]{Muzerolle1998,Muzerolle2001}. The strengths and widths of emission lines are often found to be correlated with the total accretion luminosity, albeit with a considerable scatter \citep[e.g.,][]{Natta2004,Fang2009,Rigliaco2012,Ingleby2013,Alcala2014,Alcala2017}. In cases where these correlations are verified and applicable, we can conveniently probe and measure accretion rates using emission lines alone without observing the UV-to-optical spectrum.}


So far, mass accretion rate measurements for the \pds planets have been solely based on observations of the \halpha line \citep[e.g.,][]{Wagner2018, Haffert2019, Aoyama2019, Thanathibodee2019, Hashimoto2020, Aoyama2020}. \edit1{\citet{Wagner2018} and \citet{Haffert2019} estimated the mass accretion rates by extrapolating the empirical relations between total accretion luminosity and the \halpha luminosity \citep{Rigliaco2012} or the line width \citep{Natta2004} calibrated from young stars and brown dwarfs.} However, the intrinsic uncertainties in these relations are large  and the validity of \edit1{these relations have not been observationally verified for planets}.
The planetary accretion shock models \citep{Aoyama2019,Thanathibodee2019,Szulagyi2020} all found that the \halpha emission and the accretion rate correlation for planets deviates from that for stars, due to differences in radiative transfer within the accretion shocks between protoplanets and protostars. Therefore, accurate planetary accretion rate measurements need both the \halpha line and hydrogen continuum emission. Because the \edit1{continuum} likely contains the majority of the accretion luminosity, observing and measuring \pds planets' hydrogen continuum emission are particularly imperative.



Constraining the hydrogen continuum emission requires measuring the flux density on the blue side of the Balmer jump ($\lambda=3646\,\angstrom$) in the ultraviolet (UV). In this paper, we present Hubble Space Telescope/Wide Field Camera 3 (HST/WFC3) UVIS direct-imaging observations of the \pds system in both the UV  (F336W) and the \halpha line (F656N). We adopt a novel space-based angular differential imaging (ADI, \citealt{Liu2004,Marois2006}) observational strategy to remove the contamination from the stellar point spread function (PSF) and measure the flux density of \pdsb. Using these measurements, we calculate the combined continuum$+$\halpha accretion luminosity, determine the H$\alpha{}$-to-continuum luminosity ratio, and derive the mass accretion rate for \pdsb. We then compare our measurement to previous accretion rate constraints of \pdsb, as well as to those of wide-orbit planetary-mass companions, brown dwarfs, and stars. Finally, we discuss the properties of \pdsb{}'s accretion-induced emission and its implication for the formation of this system.

\section{Observations}

We observed the \pds system using HST/WFC3 in its UVIS channel for 18 HST orbits (Program GO-15830\footnote{Detailed observing plan can be found here: \url{https://www.stsci.edu/hst/phase2-public/15830.pdf}}, PI: Zhou). All observations were conducted in the direct-imaging mode with the \texttt{UVIS2/C512C} subarray (field of view: $20.2''\times20.2''$).  The observations were constructed as six visit-sets, each consisting of three contiguous HST orbits. They were executed on UT dates starting on 2020-02-07, 2020-04-08, 2020-05-07, 2020-05-08, 2020-06-19, and 2020-07-03, respectively.  As part of our angular differential imaging (ADI) strategy, the telescope orientation angle was increased by at least ten degrees from orbit to orbit. The telescope position angle (PA) ranged from $100.8^{\circ}$ to $297.0^{\circ}$, spanning $196.2^{\circ}$ in total.

We used the F336W ($\lambda_{\mathrm{eff}}=3359$~\AA, FWHM $=550$~\AA) and F656N ($\lambda_{\mathrm{eff}}=6561$~\AA, $\mathrm{FWHM}=17.9$~\AA) filters to measure the flux on the blue side of the Balmer jump and in the \halpha line, respectively. \edit1{The F336W band flux is not significantly contaminated by any Balmer lines or Ca II H and K lines, because the filter transmission drops steeply to zero at $3700$~\AA. } Each orbit consisted of ten 120 s  F336W exposures and nine 20 s  F656N exposures. Over 18 orbits, these observations amount to integration times of 21,600 s in F336W and 3240 s in F656N.

The WFC3/UVIS native spatial resolution (pixel scale$=40$~mas, FWHM ${\sim}1.75$~pixels) is below the Nyquist limit. To improve the spatial sampling of the images, we adopted a nine-point spiral dithering pattern. The dithering step was one half pixel (20 mas). For one cycle, the telescope pointing position started at a relative origin of (0, 0) and moved along a counterclockwise spiral track ((0, 0) $\rightarrow$ (0.5, 0) $\rightarrow$ (0.5,0.5) $\rightarrow \cdots \rightarrow$ (0.5, -0.5)). At every pointing, two exposures (one in each filter) were taken. Finally, the telescope returned to the origin and took one more F336W exposure.  This dithering strategy enabled the reconstruction of images with spatial sampling of 20\,mas, better than the Nyquist criterion.

\section{Data Reduction}

We start with the \texttt{CalWFC3} pipeline-product \texttt{flc} files. These files are similar to the flat-field corrected \texttt{flt} files and have also been corrected for charge transfer efficiency losses. Our procedures include three main steps: (1) Nyquist-sampled image reconstruction; (2) primary star PSF subtraction using the Karhunen-Lo\`eve Image Processing (KLIP, \citealt{Soummer2012}) method; and (3) astrometry and photometry on the planet with PSF-subtracted images. We explain these three steps in the following subsections.

\subsection{Up-sampling Images with the Fourier Reconstruction Method}
\label{sec:up-sampling}

 We combine dithered images using the Fourier reconstruction method \citep{Lauer1999}. Compared to the \texttt{astrodrizzle} method (which produces the \texttt{drz} files available in the STScI archive), the Fourier reconstruction method guarantees Nyquist sampling and is optimized point-source PSFs. It has previously been applied in high-contrast imaging observations of exoplanets \citep{Rajan2015}. 


We implement the Fourier reconstruction method as a python-based pipeline.  We first test the pipeline using model PSFs produced by the TinyTim software \citep{Krist2011} and confirm that the difference between the reconstructed and the true PSFs is less than 1\% in intensity per pixel.  Then, we apply the pipeline to our observations. We reconstruct two images per orbit per filter. Each image is a combination of five dithered exposures. For the F336W band, the two reconstructed images are combined from exposures 1-5 and 6-10. For the F656N band, because one orbit of observations consists of nine exposures, the first exposure (for which the pointing is at the origin) is shared by both reconstructions. The size of dithering steps as executed is determined using the World Coordinate System target pointing information provided in the \texttt{fits} file header. Finally, we apply a geometric distortion correction to the reconstructed images using the solution in \citet{Bellini2011}\footnote{The band-specific solutions are only available for the broadband filters. Nevertheless, we find that the wavelength-dependent component of the correction is negligible at our interested spatial scales  (${\sim}1\arcsec$). Therefore, we use the F336W solution for both bands.}. In total, we obtain 36 up-sampled (19.8~mas/pixel image scale, FWHM=3.5 pixels) and geometrically rectified images for each band.

\subsection{Primary Star PSF Subtraction}
\label{sec:primary-subtraction}

We use the KLIP algorithm \citep{Soummer2012} to subtract the PSF of the primary star. The algorithm is performed on a $128{\times}128$ pixel ($2\farcs51\times2\farcs51$) subarray centered on \pds. First, we coalign the images by carrying out two-dimensional (2D) Gaussian fits on the PSF core of \pds and shifting the images with bi-cubic interpolation. We then select reference PSF images for each target image. To avoid self-subtraction of the astronomical signals, we impose a  minimum limit of $35^{\circ}$ difference in the telescope roll angle between the reference image and the target image. This criterion ensures that at the separation of \pdsb (170~mas) the PSFs are separated by at least 1.5 FWHMs between the target and reference images.  To avoid cross-subtraction between \pdsbc, we exclude reference images in which the PSFs of \pds c are within 1 FWHM distance of \pdsb in the target image and vice versa. After selection,  each target image has between 15 and 22 reference images.

\begin{figure*}[t]
  \centering
  \includegraphics[height=0.38\textwidth]{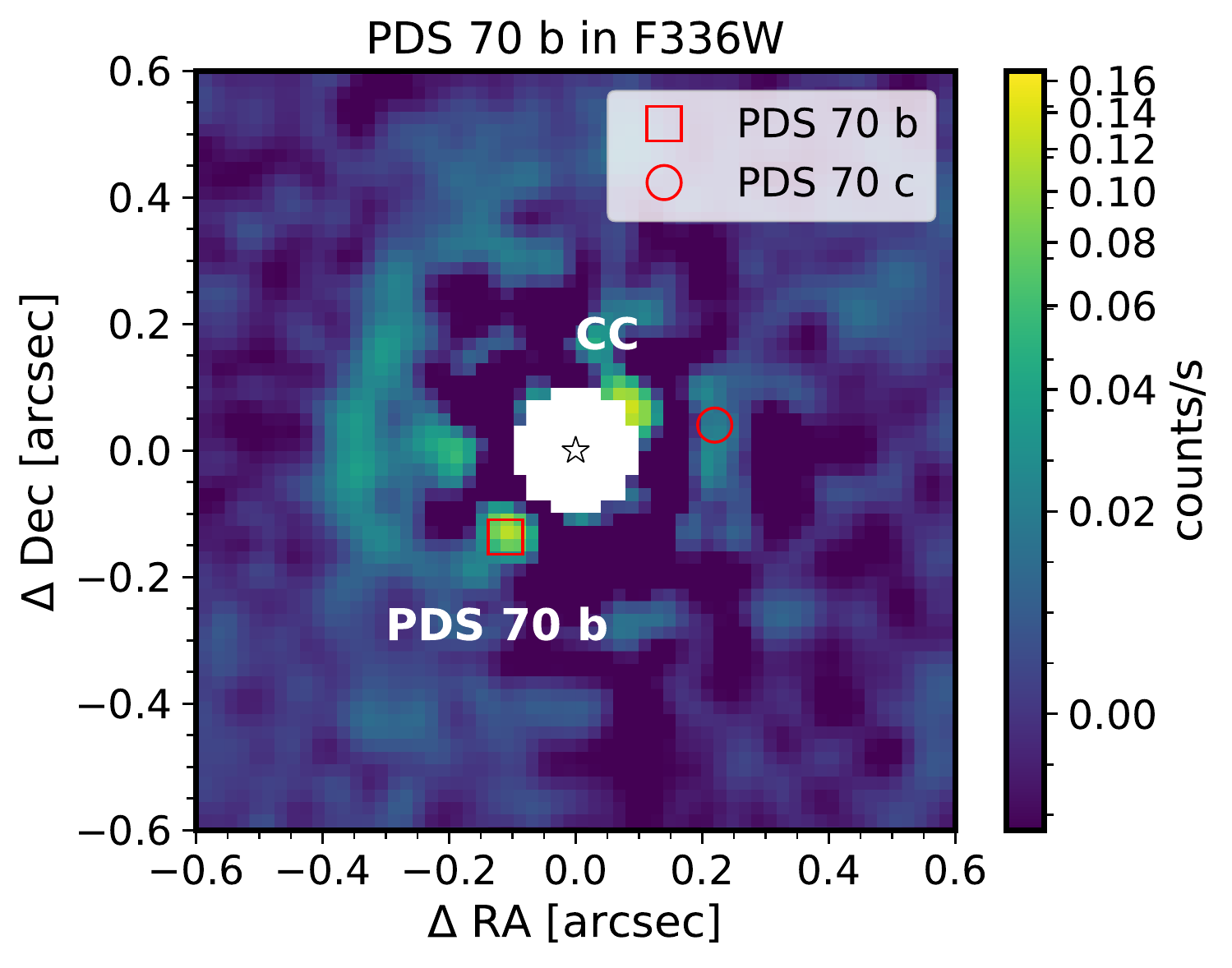}
  \includegraphics[height=0.38\textwidth]{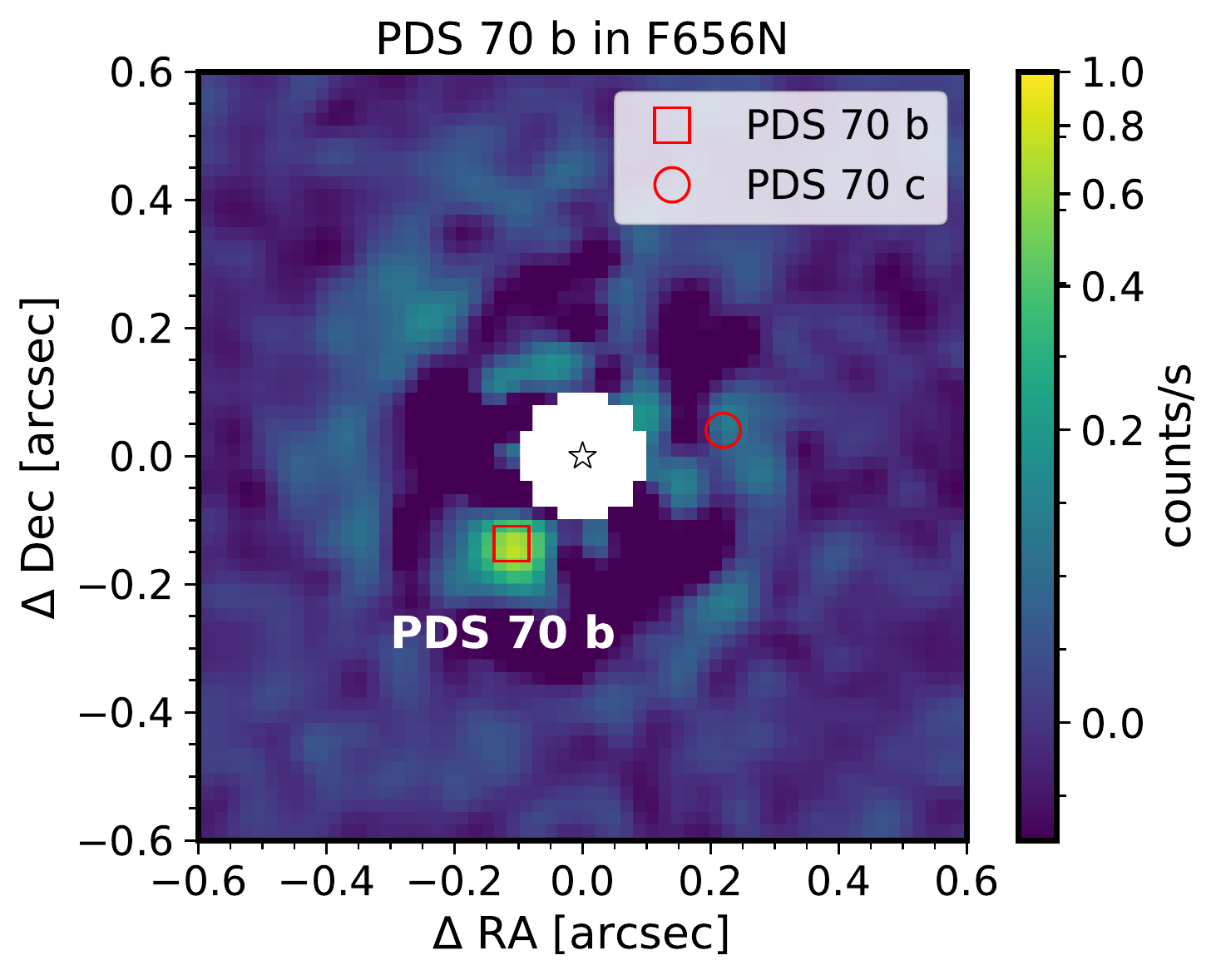}
  \caption{Primary-subtracted images in up-sampled resolution (20~mas per pixel) of the PDS70 system in the F336W (left) and F656N (right) bands. PDS70b is detected in both bands. Another close candidate companion (CC) is identified in the F336W image at $\mathrm{PA}{\sim}310$ and $\mathrm{sep}{\sim}110$~mas. To reduce high frequency noise, we smooth these images with a Gaussian kernel with a FWHM of 2.8~pixels (80\% of the PSF FWHM). The expected positions of PDS 70 b and c, which are from their most recent astrometry \citep{Wang2020}, are shown in red squares and circles, respectively. North is up and east is to the left.} \label{fig:result_images}
\end{figure*}

KLIP  is performed in parallel on subregions of the images. We experiment with various geometries for subdividing the images and select the one that results in the best planet detection SNRs. In the optimal solution, the target image is divided into annular sectors with radial bounds at 100, 350, and 700 mas. This division also defines our inner working angle to be 100~mas. Each annulus is further divided into three equal-size sectors, each spanning $120^{\circ}$.  We run separate KLIP reductions to optimize for PDS 70 b and c individually. The planet being optimized for is placed in the center of an annular sector.  After primary subtraction, we derotate the images to align their $y$-axis to the north. Finally, we register all derotated and primary-subtracted images by the centroid of PDS 70, align them using cubic interpolation shift, and combine the aligned images using the inverse-variance weighting method \citep[]{Bottom2017}.  Figure \ref{fig:result_images} shows the primary-subtracted images. Point sources that match the expected position of \pdsb are detected in both bands. \pds~c is not detected in either band.

\subsection{Photometry and Astrometry of \pdsb}
\label{sec:phot}

We conduct photometric and astrometric measurements of \pdsb using the KLIP forward modeling method \citep{Pueyo2016}. This technique calibrates measurement biases introduced in primary subtraction procedures. In our implementation, we inject the forward modeling signals into the original frame before the image reconstruction step, so it also addresses possible systematics caused by image reconstruction. We use TinyTim  PSFs \citep{Krist2011} to model the astrophysical signals. Negative-flux PSFs are injected at and around the expected position of \pdsb in the original undersampled images.  When the injected PSF matches the true signal in intensity and position, it cancels out the astrophysical signal and, therefore, locally minimizes the residuals in the primary-subtracted images. The injected PSF is sampled on a $10\times10\times10$ grid of flux density, PA, and separation. At each grid point, we run a complete data reduction and calculate the residual sum of squares (RSS) within an annular sector ($\theta=80^{\circ},r_{\mathrm{in}}=100\,\mathrm{mas}, r_{\mathrm{out}}=240\,\mathrm{mas}$, centered on \pdsb). We find that the RSS follows a parabolic function of the injected flux, PA, and separation, which is the expected outcome when the measurement biases are caused by ``over-subtraction'' instead of ``self-subtraction'' \citep{Apai2016, Pueyo2016}. We use the minimum of each best-fit parabola as the final measurements for the flux density, PA, and separation of \pdsb.

\subsection{Injection-and-Recovery Tests}
\label{sec:injection-and-recovery}
\begin{figure*}
    \centering
    \includegraphics[width=\textwidth]{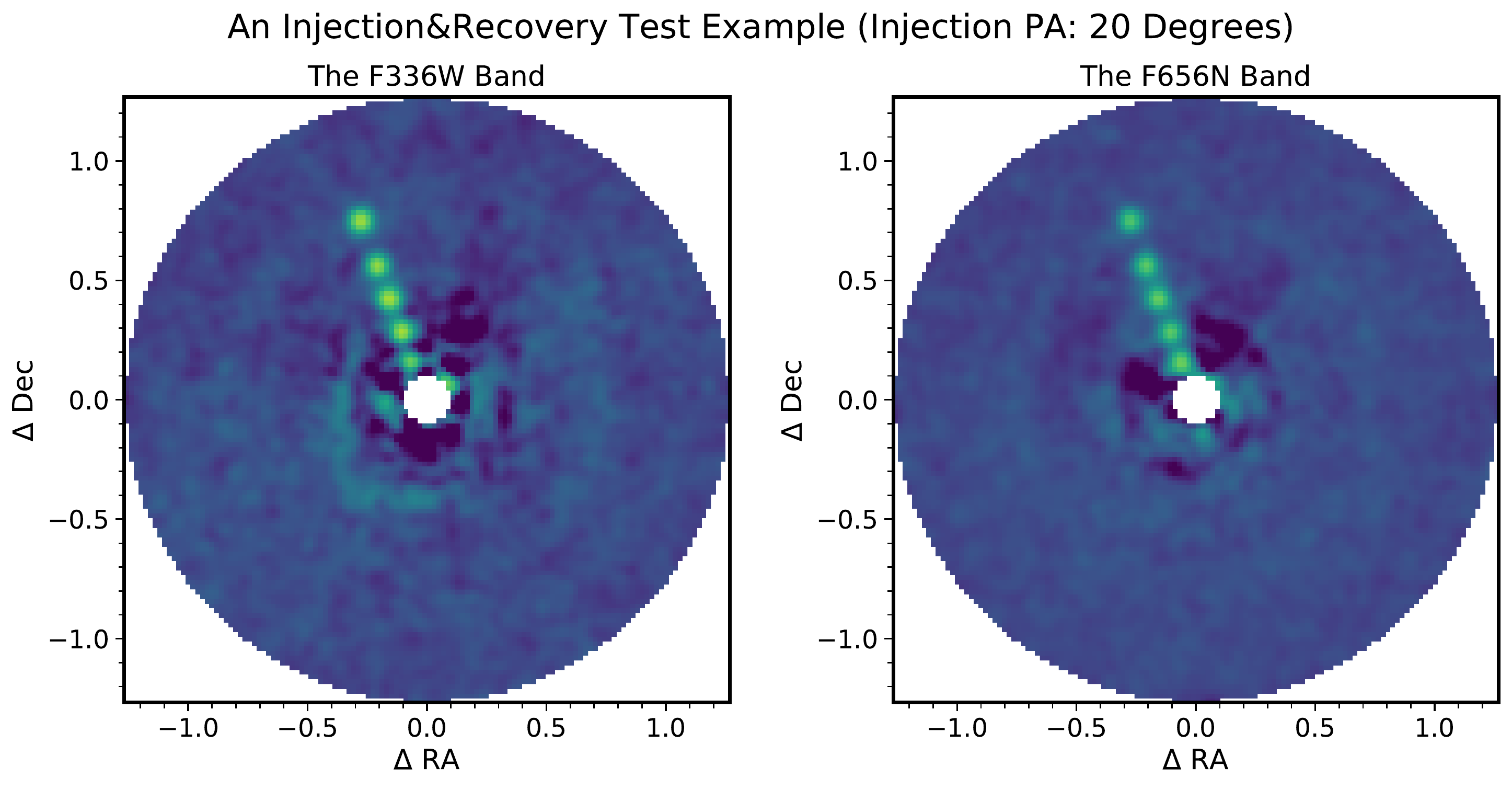}
    \caption{Examples of injection-recovery tests in the F336W (left) and F656N (right) bands. TinyTim PSFs with the same flux densities as PDS70b are injected at a PA of 20 degrees and separations at 173, 300, 450, 600, and 800\,mas. The innermost injected PSF has the same separation as PDS70b. Five injected planets are recovered in both bands. \pdsb is removed from the images by subtracting its best-fit PSF model to mitigate its interference with the test.  The same tests are repeated at another twelve position angles. }\label{fig:injection-and-recovery}
  \end{figure*}

We estimate the KLIP throughput and its uncertainty by injecting and recovering TinyTim model PSFs \citep{Krist2011} that have the same flux densities as \pdsb.  In each round of the injection-and-recovery test, we first inject five PSFs at the same PA but different separations of 173, 300, 450, 600, and 800 mas so that the innermost one has the same separation as \pdsb. We also subtract an identical PSF at the position of PDS 70 b to mitigate its interference with the test. We then perform KLIP with the same setups as those in \S\ref{sec:primary-subtraction} and conduct photometry on the primary-subtracted image (Figure~\ref{fig:injection-and-recovery}) for each injected PSF to obtain its recovered flux. The throughput factor is determined as the ratio between the recovered and the injected flux. We repeat this test for thirteen evenly spaced PAs that differ from the PA of PDS 70 b for at least $60^{\circ}$.
We take the averages of the thirteen iterations as the azimuthal-averaged throughputs and the standard deviations as the throughput uncertainties. Figure~\ref{fig:injection-and-recovery} shows examples of primary-subtracted images that contain injected PSFs.


\subsection{Signal-to-noise Ratio and Uncertainty Analyses}
\label{sec:SNR}

\begin{figure*}
    \centering
    \includegraphics[height=0.292\textwidth]{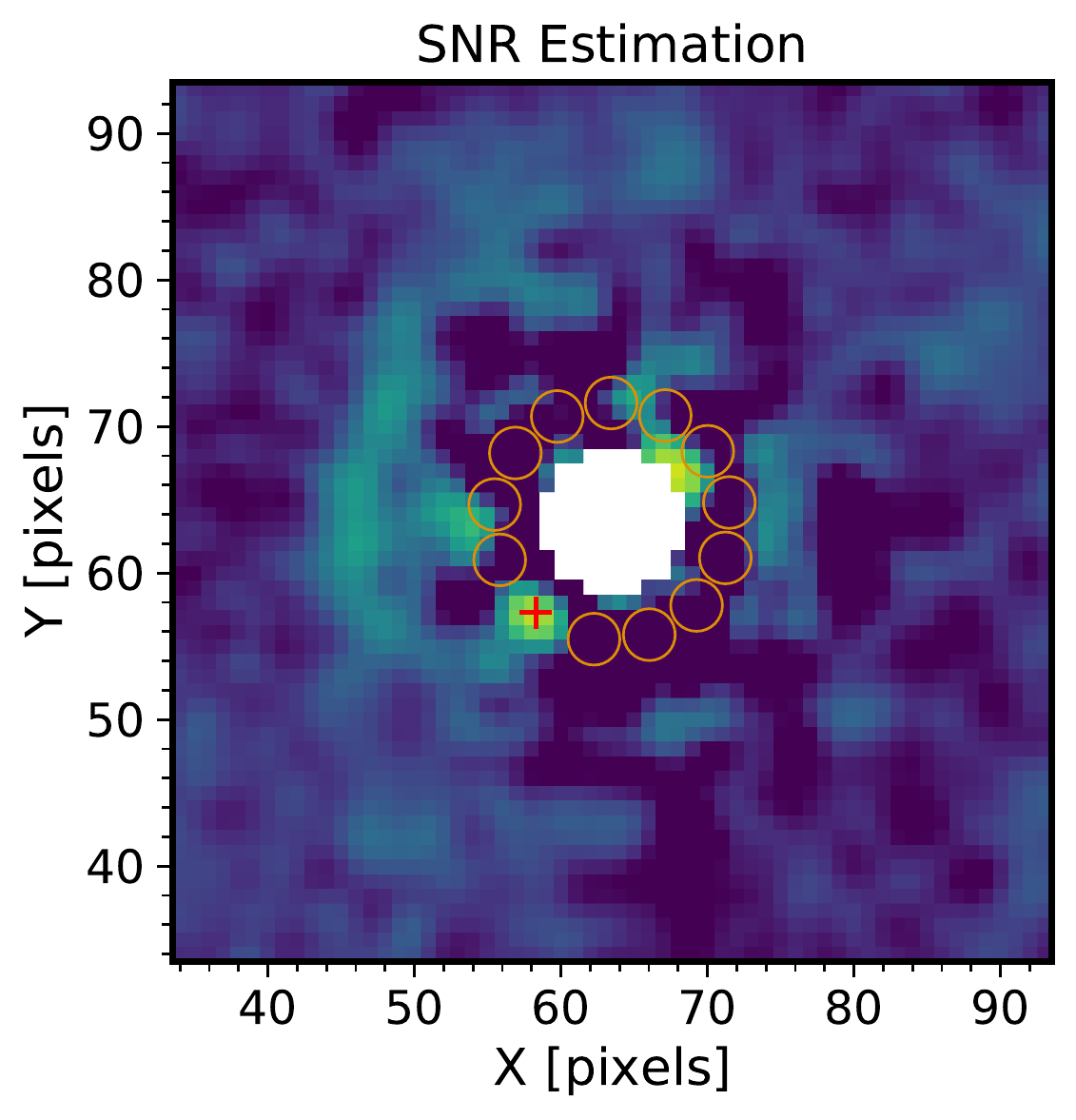}
    \includegraphics[height=0.292\textwidth]{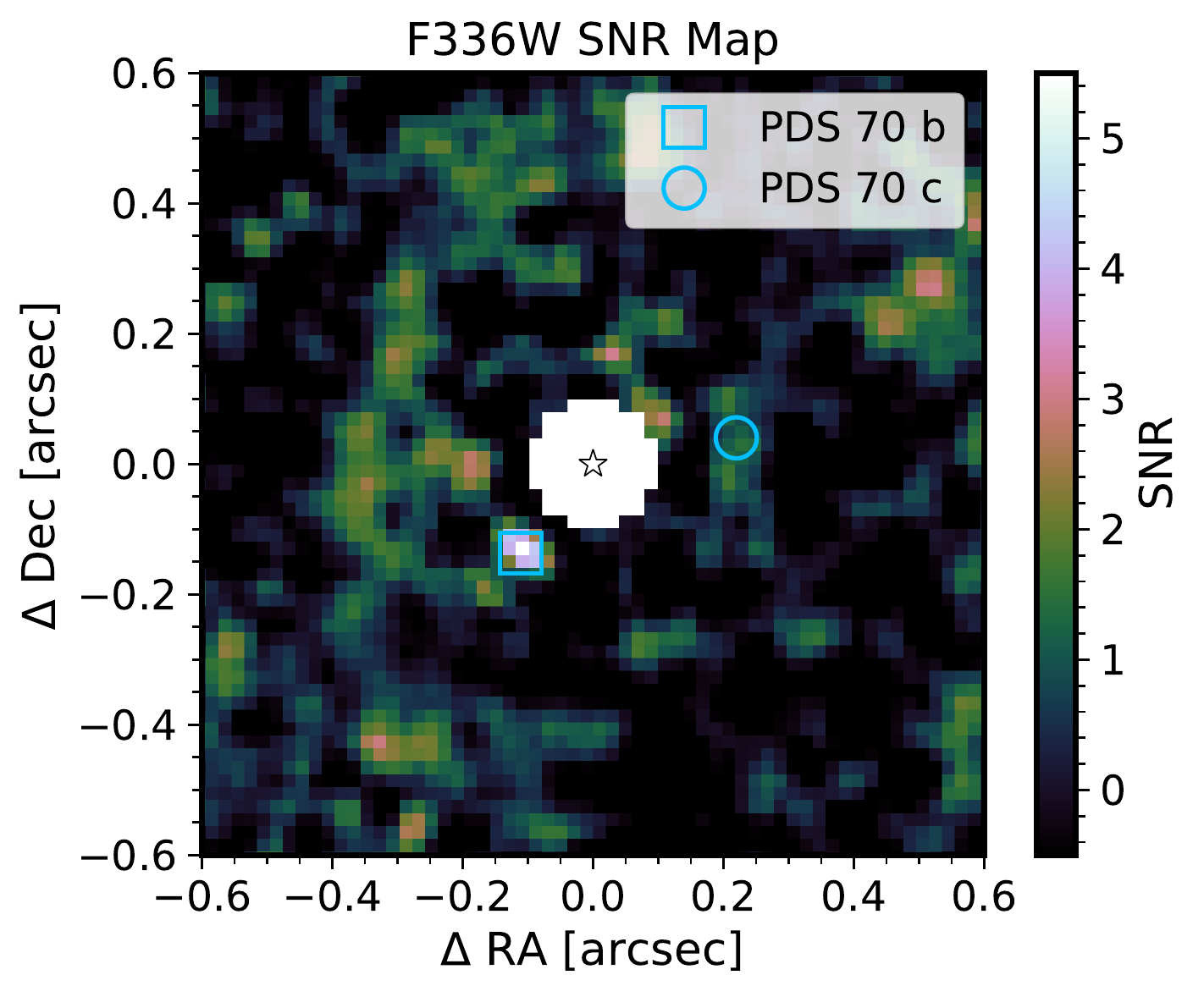}
    \includegraphics[height=0.292\textwidth]{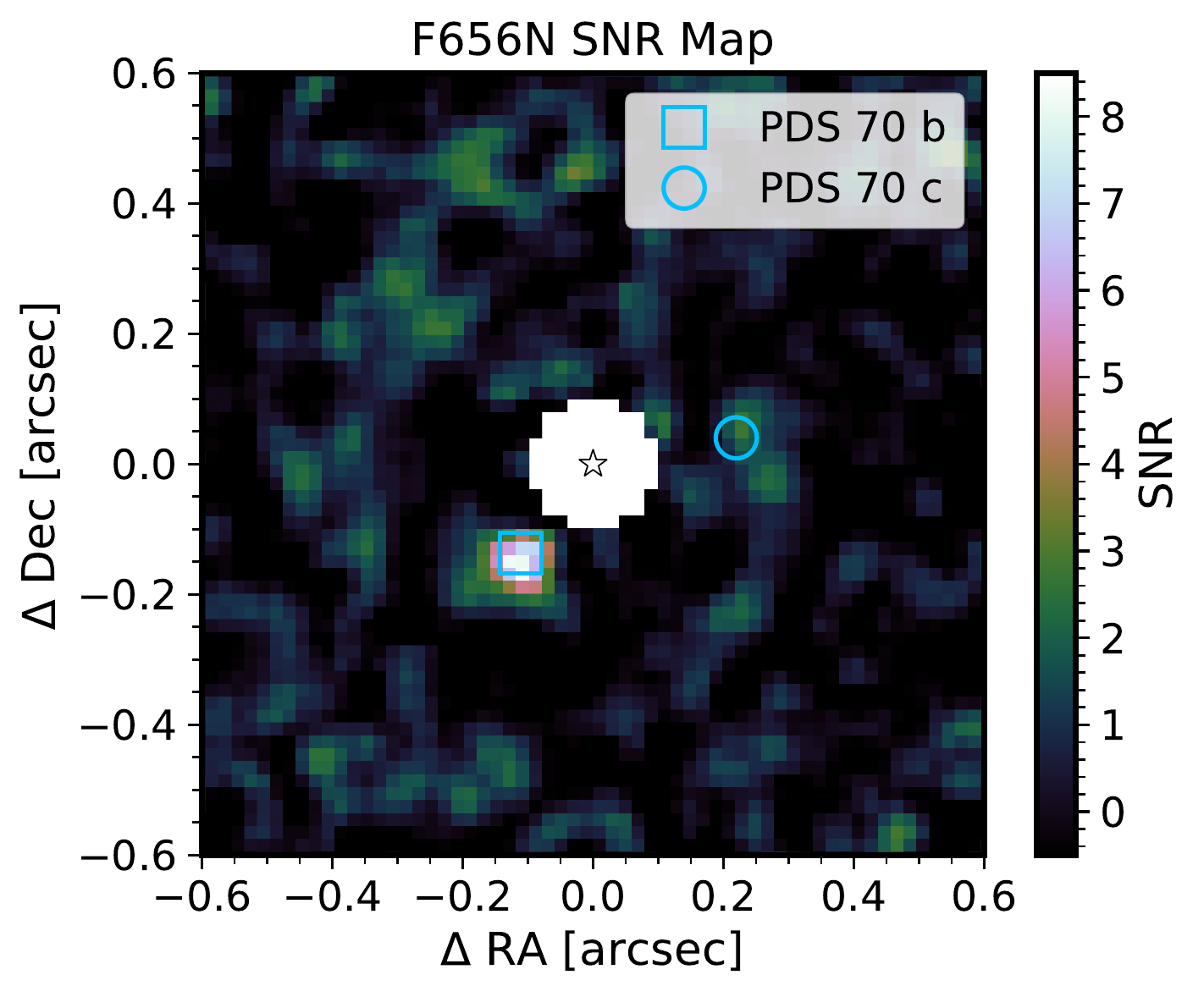}
    \caption{\edit1{Derivation of SNR and the SNR maps in the F336W and F656N bands. \emph{Left:} a demonstration of how the SNR is calculated. The red cross is the centroid position of the \pdsb detection. The orange circles represent the non-overlapping 1-FWHM diameter apertures for background and speckle noise estimation. \emph{Middle and right:} SNR maps in the F336W and F656N bands. Positions of PDS~70 b and c are marked by cyan squares and circles, respectively. The detection of \pdsb is the only significant detection in both bands.}  }\label{fig:snr_demonstration}
  \end{figure*}

We calculate the planet detection SNRs following the procedures described in \citet{Mawet2014}. This method properly accounts for small number statistics at close separations in high-contrast imaging data. First, we integrate the planetary signals in the primary-subtracted images using a 1.0~FWHM (3.5 pixels) diameter aperture. We position the aperture centered on the location of \pdsb as determined by a 2D Gaussian fit.  We then estimate the background level and noise.  Aperture-integrated fluxes are taken in non-overlapping 1.0~FWHM diameter apertures located and distributed azimuthally at the same separation as the detected point source (Figure~\ref{fig:snr_demonstration}). We insert the mean and standard deviations of these flux values into Equation (9) of \citet{Mawet2014} to derive the detection SNRs.

\edit1{SNR maps are derived using the same method. We replace the position of \pdsb by the coordinates of each pixel and repeat the SNR calculations. Iterations over the entire images result in the SNR maps. As shown in Figure~\ref{fig:snr_demonstration}, \pdsb is the only significant detection in both bands.}

  The photometric uncertainty consists of three components:  the speckle noise, the photon noise, and the KLIP throughput uncertainty. Speckle noise is determined from the standard deviation of the flux integrated within the apertures illustrated in Figure \ref{fig:snr_demonstration}. This component also accounts for possible contamination from the circumstellar disk. Photon noise is the square root of the total number of photons collected over the 18 orbits of observations. The KLIP throughput uncertainty is derived in \S\ref{sec:injection-and-recovery}. For both bands, speckle noise is the dominant component of the total error budget. It is more than ten times greater than the photon noise and a few times greater than the KLIP throughput uncertainty. We assume that these three components are independent and combine them in quadrature.  For astrometric error estimation, we assume the radial and tangential directions are independent and calculate their uncertainties as FWHM/SNR.

  \section{Results}
  
\subsection{Flux densities and the position of PDS 70 b}
\newcommand{\ElectronPerS}{\ensuremath{\mathrm{e^{-}\,s^{-1}}}\xspace}
\newcommand{\flamunit}{\ensuremath{\mathrm{erg\, s^{-1}\,cm^{-2}\,\angstrom^{-1}}}}
We detect \pdsb in the  F336W and F656N bands with aperture-integrated SNRs of 5.3 and 7.9, respectively. These SNRs correspond to false positive probabilities of $9.4\times10^{-5}$ and $8.1\times10^{-7}$ based on Student's $t$-test statistics \citep{Mawet2014}. The average planet-to-star brightness contrast ratios are $3.25\pm0.66\times10^{-4}$ and $1.38\pm0.19\times10^{-3}$ in the two bands. Neither band yields an $\mathrm{SNR}>2$ detection for \pds~c (see \S\ref{sec:pdsc}).

Our photometry for \pdsb yields count rates $1.12\pm0.22$~\ElectronPerS and $5.70\pm0.79$~\ElectronPerS in the F336W and the F656N bands, respectively. We convert count rates to flux densities ($f_{\lambda}$ in \flamunit) {using} the \texttt{PHOTFLAM} inverse sensitivity factors provided in the \texttt{fits} file headers. This results in $f_{\lambda,\,\mathrm{F336W}}=1.4\pm0.3\times10^{-18}\mathrm{erg\, s^{-1}cm^{-2}\angstrom^{-1}}$ and $f_{\lambda,\, \mathrm{F656N}}=9.2\pm1.3\times10^{-17}\mathrm{erg\, s^{-1}cm^{-2}\angstrom^{-1}}$. We derive \pdsb's \halpha{} line flux by multiplying $f_{\lambda,\, \mathrm{F656N}}$ by the effective bandpass of the F656N filter ($17.65\,\angstrom$), and find it to be $\mathcal{F_{\mathrm{F656N}}}=1.62\pm0.22 \times10^{-15}\mathrm{erg\, s^{-1}cm^{-2}}$.

\begin{figure}
    \centering
    \includegraphics[width=\columnwidth]{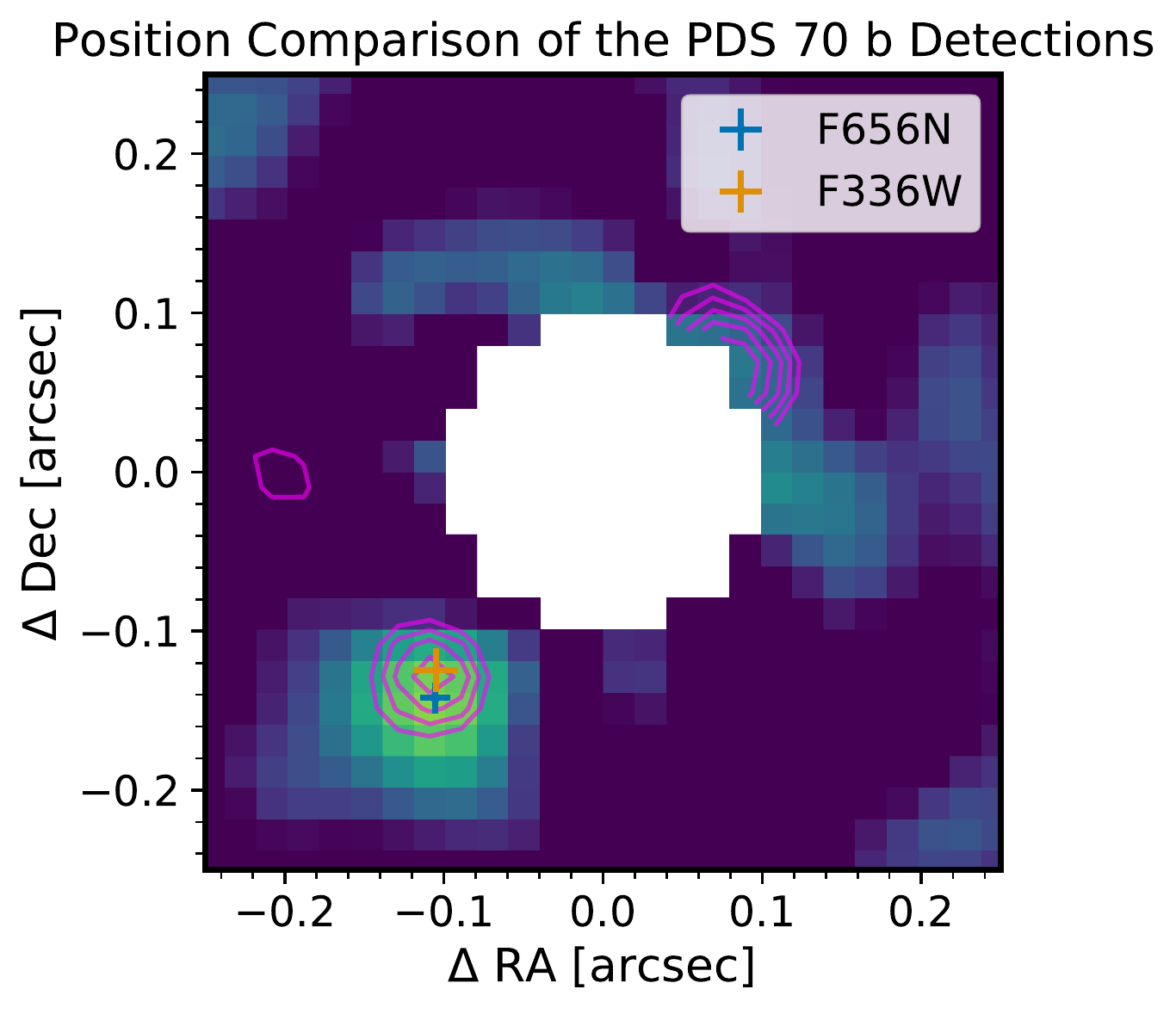}
    \caption{A comparison of the positions of the point sources detected in the F336W and F656N bands. Contours (30\%, 40\%, 50\%, 60\%, and 70\% of the maximum pixel value) of the F336W image are overlaid on the F656N image. The F336W contours match the position of the \pdsb's detection in F656N. The two crosses mark the best-fit positions of PDS70 b in the two bands. The sizes of the crosses represent the $1\sigma$ uncertainties.  The proximity of the two crosses indicates that \pdsb's detections in the two bands are consistent with each other in their positions.}\label{fig:astrometry_comparison}
  \end{figure}

We determine \pdsb's  position angles and separations in the two bands separately. The results are $\mathrm{PA}=139.9\pm5.0^{\circ}$, $\mathrm{sep}=163\pm14$~mas in F336W  and $\mathrm{PA}=143.4\pm3.0^{\circ}$, $\mathrm{sep}=177.0\pm9.4$~mas in F656N.  The uncertainties in the F336W band are greater due to the lower detection SNR. As demonstrated in Figure~\ref{fig:astrometry_comparison}, \pdsb's positions in the two bands are consistent within $1\sigma$ .

  We also identify excess emission of a close candidate companion (CC) at $\mathrm{PA}{\sim}310^{\circ}$ and $\mathrm{sep}{\sim}110$~mas in the F336W image. The location is close to the inner working angle (100~mas). The detection SNR is ${\sim}2.5$, corresponding to a false positive probability of 1.8\%. \citet{Mesa2019a} reported a point-like feature at a similar location in their VLT/SPHERE observations. Based on its near-infrared spectrum, \citet{Mesa2019a} interpreted it as scattered starlight from circumstellar material.

  \subsection{Contrast Curves}

  \begin{figure*}[t!]
    \centering
    \includegraphics[width=0.48\textwidth]{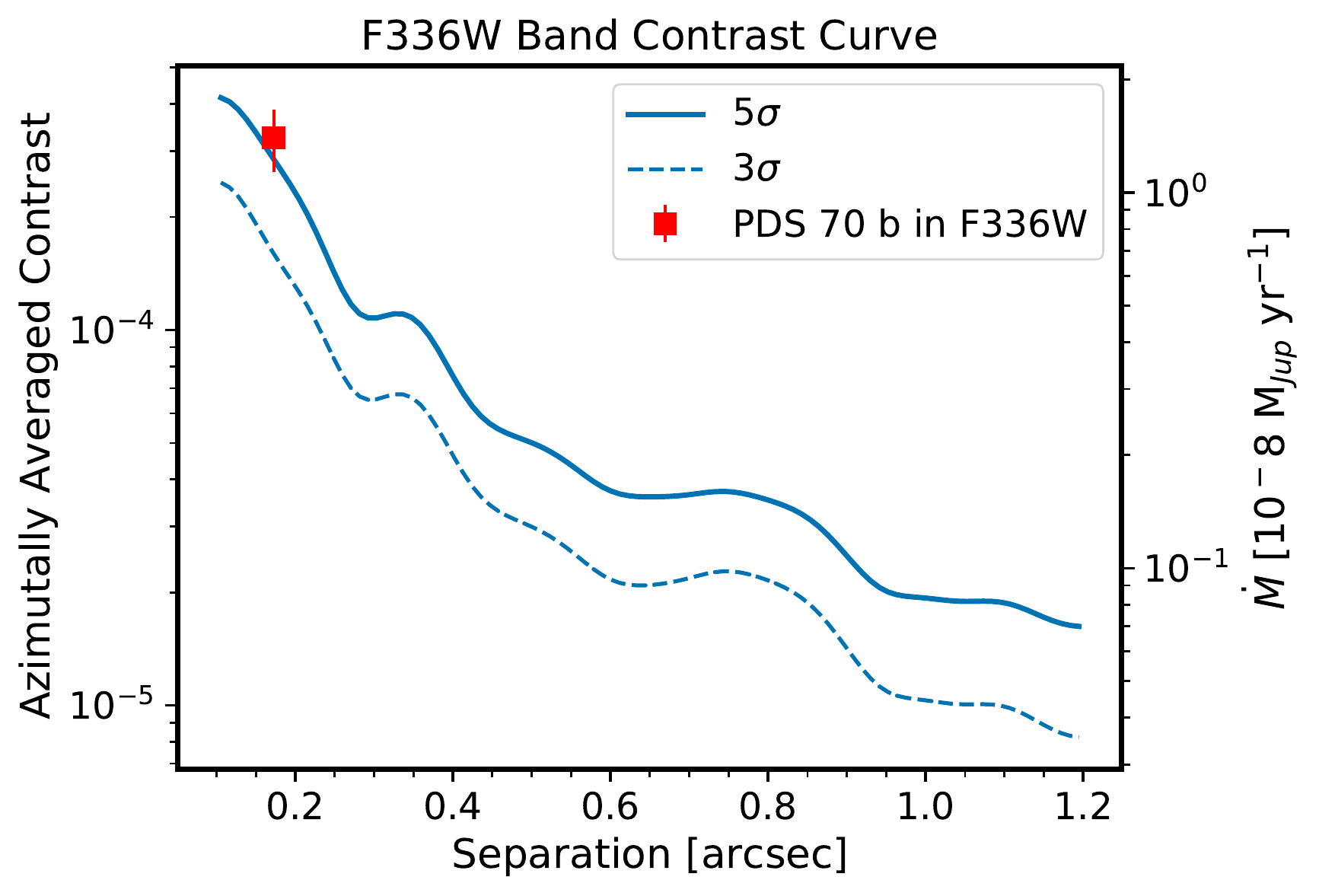}
    \includegraphics[width=0.48\textwidth]{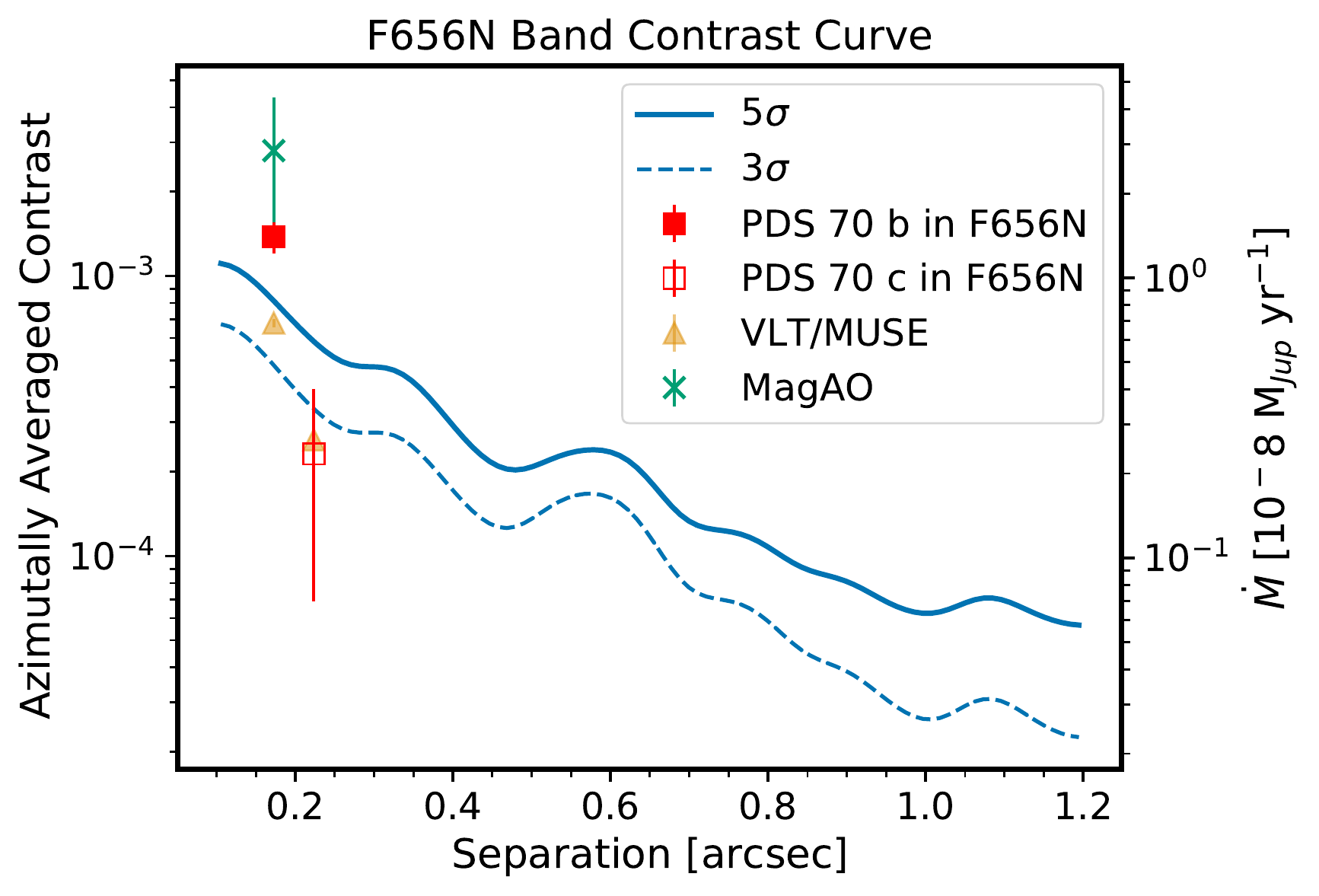}
    \caption{Azimuthally averaged contrast curves in the F336W and F656N bands. Blue solid and dashed lines represent the  $5\sigma$ and $3\sigma$ contrast curves, respectively. The secondary $y$-axes show the corresponding sensitivity in $\dot{M}$ of a \pdsb-like planet. We convert contrast ratios to $\dot{M}$ sensitivities based on Equation~\ref{eq:mdot}. The red squares are the observed contrasts for PDS 70 b. In the right panel, the open red square marks the aperture-integrated signal at the expected position of PDS 70 c. The right panel also shows ground-based \halpha contrast measurements of PDS 70 b and c \citep{Wagner2018, Hashimoto2020}.  }\label{fig:contrast-curve}
  \end{figure*}

The contrast curves, defined as the flux ratio of the detection limit and  \pds, are presented in Figure \ref{fig:contrast-curve}.  We use the method of \S\ref{sec:SNR} to estimate the $3.5\,\mathrm{pixels}$ diameter aperture-integrated detection limit and then apply aperture and throughput corrections to account for the finite aperture size and over-subtraction. Aperture correction coefficients are determined by interpolating the WFC3/UVIS2 encircled energy table\footnote{See \url{https://www.stsci.edu/hst/instrumentation/wfc3/data-analysis/photometric-calibration/uvis-encircled-energy}.}. Throughput calibration factors are derived with the injection-and-recovery tests (see \S\ref{sec:injection-and-recovery}). For separations that are not  in the injection-and-recovery tests, the linearly interpolated (or extrapolated) values are used. The \pds's flux densities are the time-averaged values over the entire observations. Based on the contrast curves, the F336W image is more sensitive than the F656N by a factor of 3--4, depending on the separation. The 5$\sigma$ detection limit in F336W reaches a contrast of $1.1\times10^{-4}$ at {0.3~arcsec} and $2.0\times10^{-5}$ at 1~arcsec. In the F656N band, these limits are $4.6\times10^{-4}$ and $6.8\times10^{-5}$, respectively.

For comparison, we overplot the observed contrasts of \pdsb in F336W and F656N, as well as the \halpha contrasts of \pdsbc measured in MagAO \citep{Wagner2018} and VLT/MUSE \citep{Haffert2019, Hashimoto2020} observations. To account for the filter bandpass and spectral resolution differences, we calibrate these \halpha contrasts and unify them under the WFC3/UVIS2/F656N system.  We adopt the planets' absolute \halpha flux \citep{Wagner2018, Hashimoto2020} and divide them by the F656N band-integrated flux of \pds ($\mathcal{F}_{\mathrm{PDS\,70,\, F656N}}=1.18\times10^{{-12}}\,\mathrm{erg\,cm^{-2}s^{-1}}$). These observations demonstrate clear disagreement in \pdsb's \halpha contrasts. As for PDS 70 c, its \halpha contrast estimated by VLT/MUSE is below our $3\sigma$ sensitivity limit. 


\subsection{Time-resolved Photometry of PDS 70 b in \halpha{}}
\label{sec:time_resolved}
  
\begin{figure*}
    \centering
  \includegraphics[width=\textwidth]{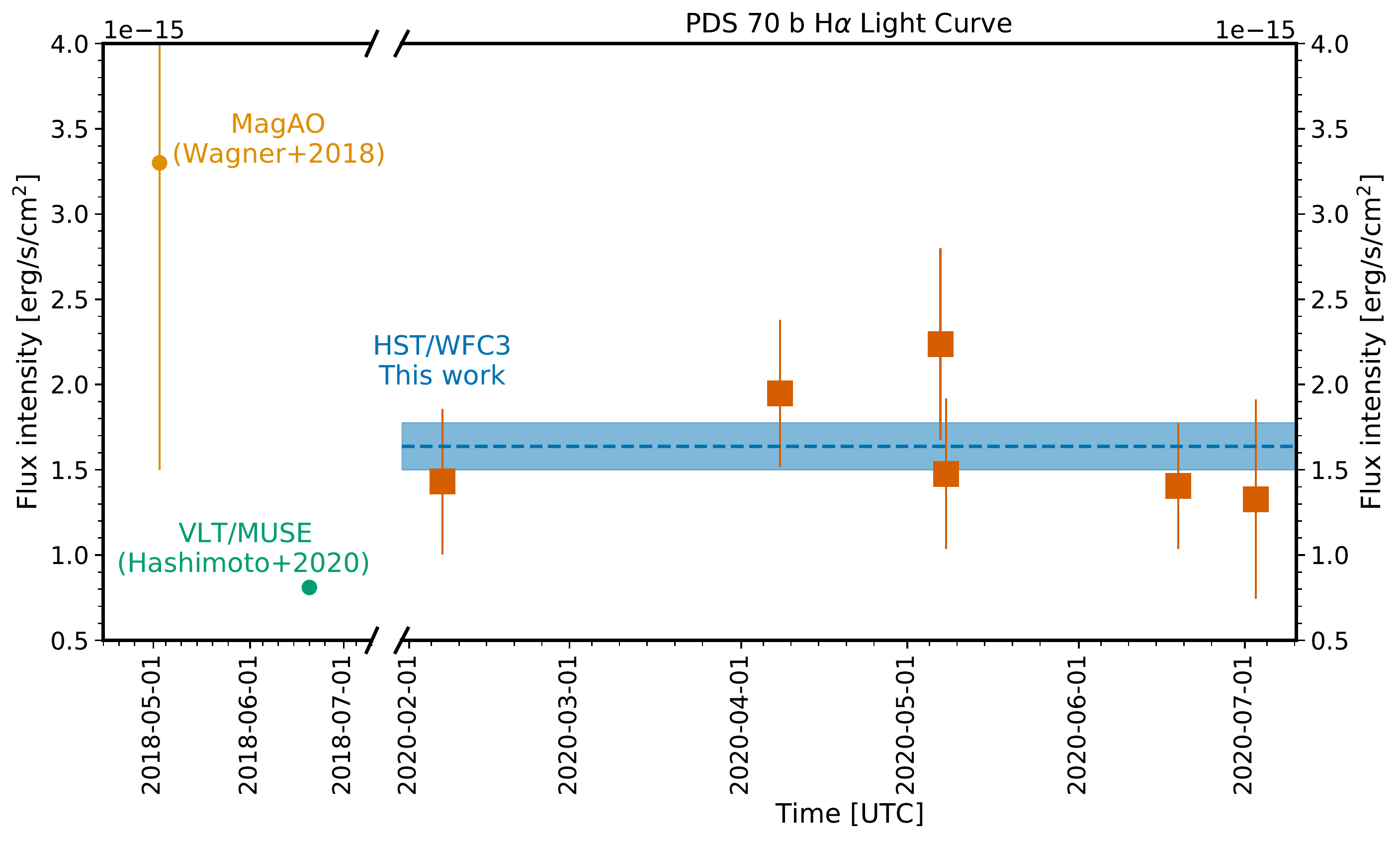}
    \caption{Time-resolved \halpha flux of PDS70b. \edit1{Yellow and green circles and associated errorbars (1$\sigma$) show two previous ground-based measurements by MagAO \citep{Wagner2018} and VLT/MUSE \citep{Hashimoto2020}, respectively. The ground-based observations were taken ${\sim}20$ months prior to the HST data.} Red squares and errorbars are PDS70 b's \halpha flux in individual visit-sets. The blue dashed line and rectangle shade show the average HST flux and its $1\sigma$ uncertainty, respectively. \edit1{The HST measurements do not support large amplitude ($>30\%$) \halpha variability during the six epochs spanning five months.}} \label{fig:lightcurve}
  \end{figure*}

  All F656N images from the six individual visit-sets yield SNR$>3$ detections of \pdsb. Following the procedures of \S\ref{sec:phot} and \ref{sec:SNR}, \edit1{we measured \pdsb's \halpha{} flux in each visit-set to form a sparsely sampled light curve (Figure~\ref{fig:lightcurve}). This light curve does not show evidence for variability. } The average flux is within $1\sigma$ uncertainty of every time-resolved measurement except Visit-set 3, which has the strongest \halpha flux. However, the \halpha{}  flux in Visit-set 3 is only  $1.4\sigma$ greater than the average of the rest of the observations. Based on {the uncertainty of the light curve}, we place an upper limit of $30\%$ variability for \pdsb in \halpha{} \edit1{in the six epochs over a five months baseline.}

  We also compare \pdsb's \halpha flux in our HST observations to ground-based measurements from MagAO \citep{Wagner2018}, and VLT/MUSE \citep{Haffert2019, Hashimoto2020}. The HST measurement of $1.62\pm0.23\times10^{-15}\,\mathrm{erg\,s^{-1}\,cm^{-2}}$ is lower than the MagAO result of $3.3\pm1.8\times10^{-15}\,\mathrm{erg\,s^{-1}\,cm^{-2}}$ \citep{Wagner2018}  but the difference is within $1\sigma$. Our result is higher than the VLT/MUSE flux of $0.81\pm0.03\times10^{-15}\,\mathrm{erg\,s^{-1}\,cm^{-2}}$, \citep{Hashimoto2020} by $3.5\sigma$, which may suggest significant \halpha variability on ${\sim}1$-2 years timescales.
  \edit1{We note that HST and ground-based observations differ in instruments, spectral transmissions, ADI setups, and post-processing procedures. In each step, consistency in flux calibrations needs to be maintained to eliminate possible systematic error when comparing photometry. Therefore, accurately cross-calibrating these measurements is challenging and beyond the scope of this study. }
   Follow-up observations with the same instrument and consistent flux calibrations are necessary to \edit1{further evaluate whether \pdsb is variable beyond the consistent \halpha flux measured in our six epochs of observations. }

  \subsection{The Nondetection of PDS 70 c}\label{sec:pdsc}

  Neither the F336W nor the F656N image yields an $\mathrm{SNR}>2$ detection for \pds~c. The nondetection of \pds~c in the F656N band is consistent with its \halpha flux measurements reported by \citet{Haffert2019} and \citet{Hashimoto2020}. Using the method of \S\ref{sec:SNR}, we calculate the aperture-integrated (diameter=3.5~pixels) SNR at PDS 70 c's expected position ($\mathrm{PA}{=}280^{\circ}$, $\mathrm{sep}{=}223$~mas, \citealt{Wang2020})  in the F656N image and obtain a result of $\mathrm{SNR}{=}1.4$ (false positive probability of 9.5\%). The corresponding aperture and throughput-corrected \halpha flux is $2.6\pm1.9\times10^{-16}\mathrm{erg\,s^{-1}cm^{{-2}}}$, consistent with the VLT/MUSE measurement of $3.1\pm0.3\times10^{-16}\mathrm{erg\,s^{-1}cm^{{-2}}}$ within $1\sigma$. However, such a low flux does not permit a statistically significant detection in our observations. As shown in Figure~\ref{fig:contrast-curve}, both our observed \halpha flux at the position of \pds~c and the literature value are below the 3$\sigma$ detection limit. As for the F336W band,  we expect the planet is even less likely to be detected due to greater star-to-planet brightness contrast. Therefore, these nondetections are most likely due to our sensitivity limits.

  \section{Accretion onto PDS 70 \lowercase{b}}
  \newcommand{\mdot}{\ensuremath{\dot{M}}\xspace}
  
  \subsection{Estimating the Accretion Luminosity and Mass Accretion Rate of PDS 70 \lowercase{b}}

  \begin{figure*}
  \centering
  \includegraphics[width=\textwidth]{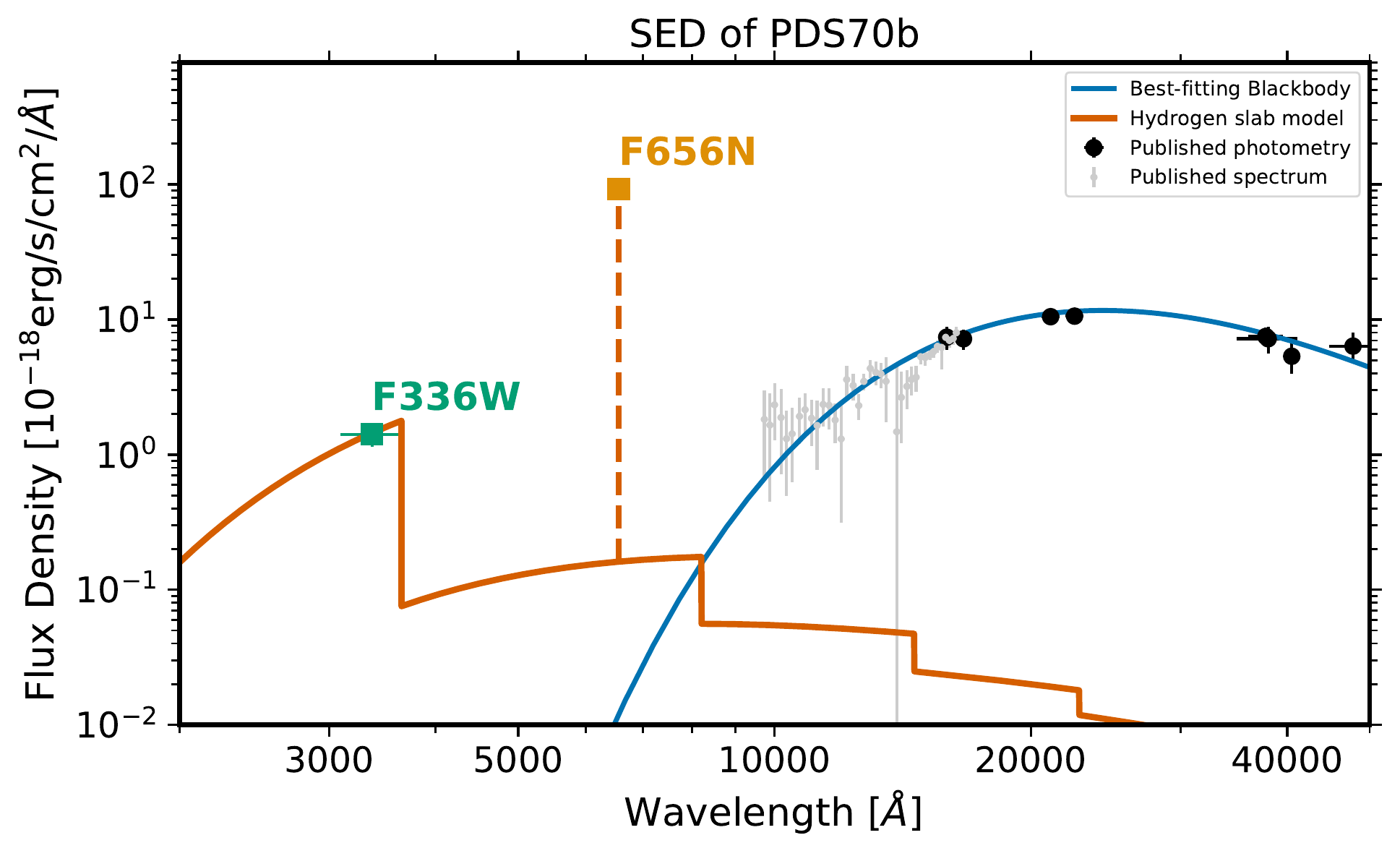}
  \caption{The UV, optical, and IR SED of \pdsb. The green and orange squares show the observed flux densities of \pdsb in the F336W and F656N bands, respectively. \edit1{We note that these are the mean flux densities within each filter bandpass.} These flux densities are more than three orders of magnitude higher than the $1.2\times10^{3}$~K blackbody (blue curve) that is the best-fit model to the planet's IR observations (gray and black dots; data taken from \citealt{Muller2018,Mesa2019a,Wang2020,Stolker2020}).  The vertical dashed line indicates the position of \halpha and the red solid curve represents a hydrogen continuum emission model spectrum. We use this model to conduct bolometric correction for the F336W photometry.}
  \label{fig:sed}
\end{figure*}

The F336W and F656N bands constrain hydrogen Balmer continuum and \halpha emission, both of which are accretion indicators \citep{Aoyama2020}. 
As shown in the UV, optical, and IR spectral energy distribution (SED) of \pdsb (Figure \ref{fig:sed}), our observed F336W and F656N flux densities are more than three orders of magnitude higher than the  best-fit {blackbody} to the infrared (IR) observations. This suggests that accretion-induced hydrogen emission dominates the flux in the F336W and F656N bands and the contribution from the blackbody continuum can be safely ignored.  Therefore, we can measure the accretion luminosity (\Lacc) with the F336W and F656N flux densities and then convert the accretion luminosity to a mass accretion rate (\mdot).  

To calculate \Lacc, we need to introduce four assumptions:
\begin{enumerate}
\item \halpha dominates the hydrogen line emission.
\item We adopt a plane parallel pure hydrogen slab model \citep{Valenti1993} to conduct a bolometric correction for the F336W flux density and derive \Lcont. This model calculates the hydrogen bound-free emission spectrum as a function of temperature, number density ($n$), slab length, and the filling factor. \citet{Herczeg2008}, \citet{Herczeg2009}, and \citet{Zhou2014} used the same model for their  bolometric corrections.
\item The hydrogen number density of the slab is $10^{13}\mathrm{cm^{-3}}$. This number density is consistent with the estimates in \citet{Aoyama2019} and \citet{Hashimoto2020} and leads to a relatively large Balmer jump (a flux ratio of 17) compared to those typically observed in classical T Tauri stars \citep[flux ratio $<3$; e.g.,][]{Valenti1993,Calvet1998,Alcala2014}, \edit1{although some brown dwarfs and a few stars have larger Balmer jumps \citep{Herczeg2009,Rigliaco2012,Alcala2014}}.
  \item We assume no extinction for \pdsb and  do not de-redden the observed flux densities.
  \end{enumerate}
  We first use these assumptions to derive \pdsb's \Lacc and then discuss their impact on the results.

\Lacc is the sum of two parts: the hydrogen line emission luminosity (\Lline) and the hydrogen continuum emission luminosity (\Lcont):
\begin{equation}
  \label{eq:1}
  \Lacc = \Lline + \Lcont.
\end{equation}
To estimate the line emission flux, we apply Assumption (1) to get  \Lline=\Lhalpha. To calculate the continuum emission flux, we conduct a bolometric correction to the F336W flux density using the hydrogen continuum spectrum model determined by Assumption (2) and (3). The continuum model spectrum is scaled so that it produces the observed F336W flux density.  Integrating the scaled model over its entire spectral range results in the continuum flux. We convert flux to luminosity by multiplying by $4\pi d^{2}$, where $d$ is \pds's Gaia DR2 distance \citep[113.06~pc,][]{Brown2016,Gaia-Collaboration2018, Bailer-Jones2018}. This gives $\Lline=6.5\pm0.9\times10^{-7}\Lsol$, $\Lcont=1.2\pm0.2\times10^{-6}\Lsol$, and $\Lacc=1.8\pm0.2\times10^{-6}\Lsol$.

We assume a magnetospheric accretion model to convert accretion luminosity to mass accretion rate (\mdot).  Following \citet{Gullbring1998}, we have
\begin{equation}
  \label{eq:mdot}
  \mdot =\Bigl(  1-\frac{R_{\mathrm{p}}}{R_{\mathrm{in}}}\Bigr)\frac{R_{\mathrm{p}}}{GM_{\mathrm{p}}}\Lacc.
\end{equation}
The accretion flow is launched at the magneto-truncation radius of $R_{\mathrm{in}}$  and free-falls onto the planetary surface. The entire kinetic energy of the accretion flow is converted to accretion luminosity. We follow \citet{Gullbring1998} and assume $R_{\mathrm{in}}=5R_{\mathrm{p}}$ \edit1{for consistency with accretion rate measurements for stars, despite the uncertainty of this geometry for the planetary regime. } Inserting our measured value of $\Lacc$ ($1.8\pm0.2\times10^{-6}\Lsol$) and the latest mass ($1M_{\mathrm{Jup}}$) and radius ($1.75R_{\mathrm{Jup}}$) estimates \citep{Stolker2020}, we get
\begin{equation}
  \label{eq:result}
  \mdot=1.4\pm0.2 \times10^{-8}\Bigl( \frac{R_{\mathrm{p}}}{1.75R_{\mathrm{Jup}}} \Bigr) \Bigl( \frac{M_{\mathrm{p}}}{1M_{\mathrm{Jup}}}
  \Bigr)^{-1}M_{\mathrm{Jup}}\,\mathrm{yr^{-1}}.
\end{equation}
The quoted uncertainty in $\mdot$ contains those propagated from $\Lacc$ but not systematic uncertainties of $\mdot$.
\edit1{The impact of assumptions and systematic uncertainties are discussed in \S\ref{sec:sys}.}

Equation~\ref{eq:mdot} provides a correspondence between the contrast ratio (Figure~\ref{fig:contrast-curve}) and \mdot of a \pdsb-like planet. Because accretion excess emission dominates the planet's flux in the F336W and F656N bands and \mdot scales linearly with \Lacc (when constant mass and radius are assumed), the contrast ratio should also scale linearly with \mdot. We conduct the contrast ratio to \mdot conversion and show the results as the secondary $y$-axes of Figure~\ref{fig:contrast-curve}. At $1\arcsec$ separation, both bands are sensitive to $\mdot$ of $10^{-9}M_{\mathrm{Jup}}\,\mathrm{yr^{-1}}$ accretion onto a \pdsb-like planet.

\subsection{Systematic Uncertainties in the Accretion Rate Estimate}
\label{sec:sys}

  Systematic uncertainties in our accretion rate estimate for PDS 70 b are similar to those of other (sub)stellar accretion measurements made with $U$-band spectrophotometry \citep[e.g.,][]{Herczeg2008}. There are four sources of systematics: bolometric correction for the continuum luminosity, neglected emission lines, uncertain accretion mechanisms, and the unknown extinction.

  The bolometric correction is directly set by the size of the Balmer jump, which is unconstrained by our observations. For our analysis, we fix the  number density of the hydrogen slab to $1\times10^{{13}}$cm$^{-3}$, resulting in a factor of 17 flux increase at the Balmer jump. This number density is low compared to those adopted in stellar accretion rate analyses \citep[e.g.,][]{Valenti1993, Gullbring1998, Ingleby2013}, but consistent with the results from analyzing \pdsb's \halpha line profile  \citep{Hashimoto2020}. Increasing the number density by one order of magnitude will decrease the Balmer jump size to 4, increase the bolometric correction factor by 20\%, and increase the accretion rate measurement by 12\%.

  We include only H$\alpha$ for calculating \Lline. Because line emission accounts for a greater amount of accretion luminosity in the substellar regime \citep{Herczeg2009,Rigliaco2012,Zhou2014}, excluding other lines may introduce greater errors in \mdot for \pdsb compared to similar measurements in the stellar regime \citep[e.g.,][]{Herczeg2008}. Based on a planetary accretion shock model, \citet{Aoyama2018, Aoyama2020} found that the Ly$\alpha$  emission can be more than one order of magnitude more energetic than the H$\alpha$ emission in accretion onto \pdsb-like planets. In this case, the hydrogen line emission will dominate PDS 70 b's accretion excess emission.  In the case where \pdsb's \Lline is the same as its \Lcont, the accretion rate will be increased by 26\% compared to the  Equation~\ref{eq:result} result.

  By adopting Equation \ref{eq:mdot}, we assume the accretion flow is launched 5$R_{\mathrm{p}}$ away from the planet and hits the planetary surface. When a different accretion paradigm is assumed \citep[e.g.,][]{Szulagyi2020}, as long as the kinetic energy of the accretion flow is entirely converted to accretion luminosity, Equation~\ref{eq:mdot} should still hold.  Based on Equation \ref{eq:mdot}, the accretion rate is correlated with the assumed radius and anticorrelated with the assumed mass. An overestimated radius or an underestimated mass will lead to an overestimated accretion rate, and \emph{vice versa}.

\begin{figure*}
    \centering
    \includegraphics[width=0.48\textwidth]{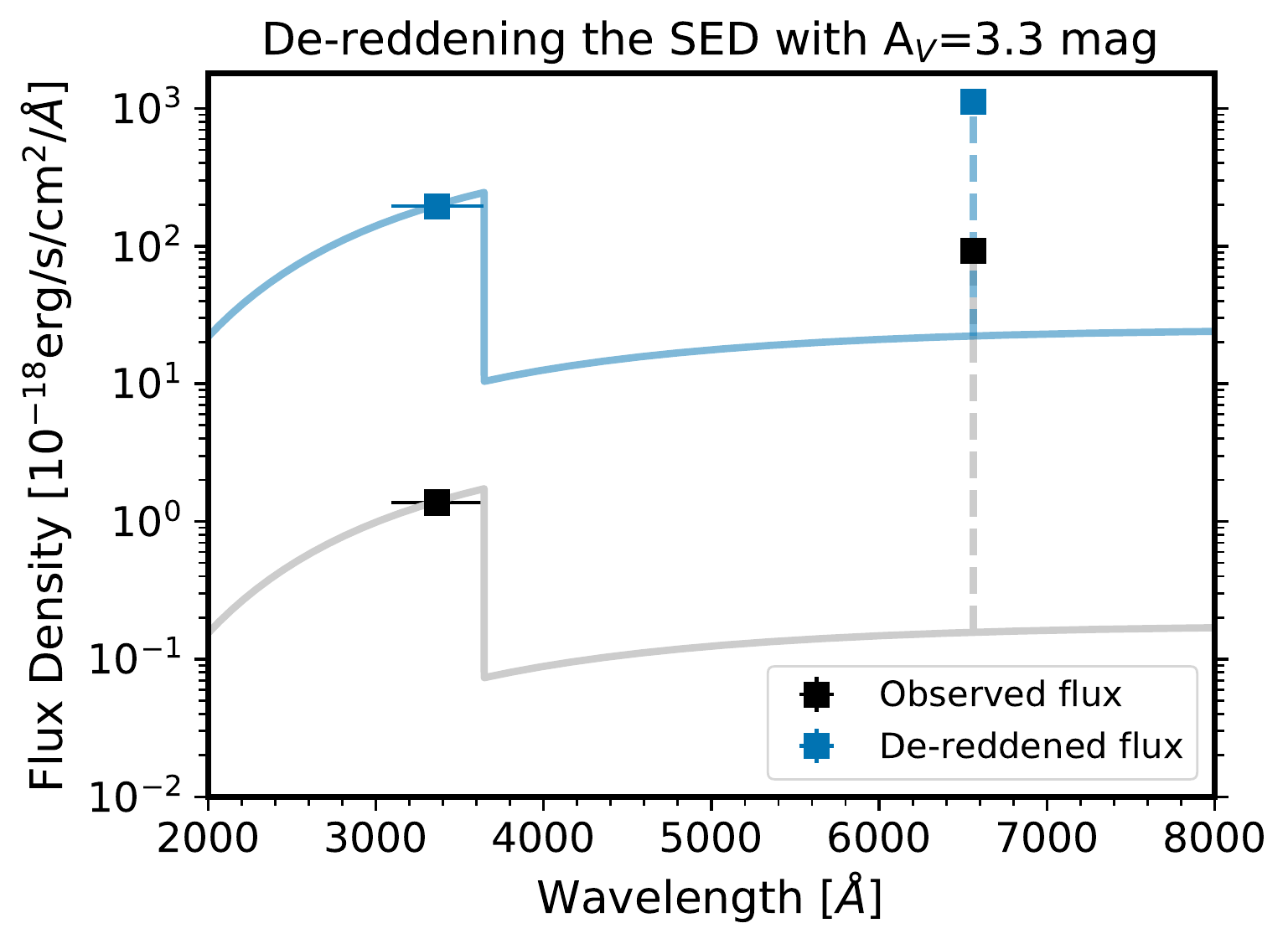}
    \includegraphics[width=0.48\textwidth]{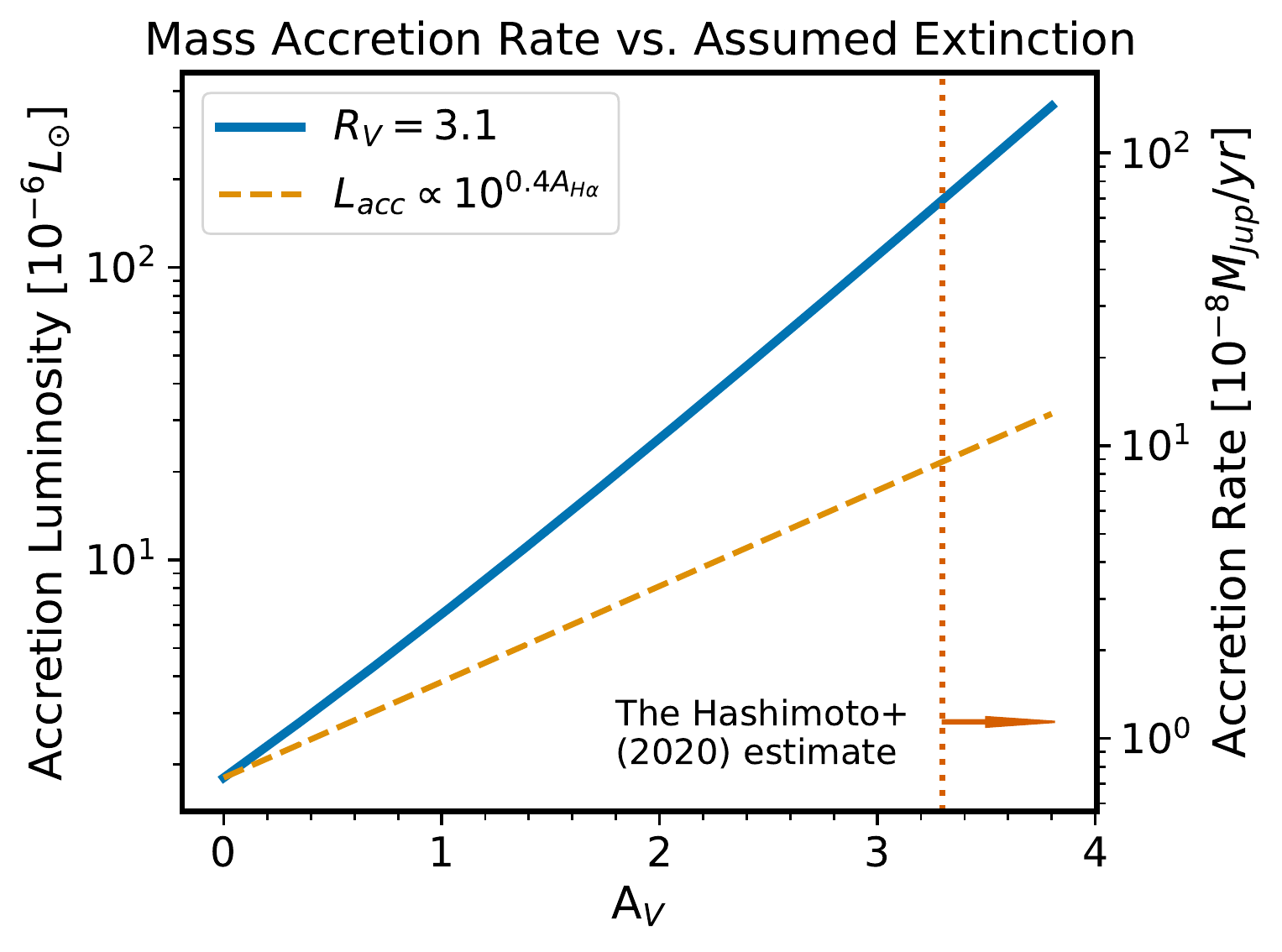}
    \caption{Extinction affecting the accretion luminosity and accretion rate measurements for \pdsb. \emph{Left:} comparison between the observed flux densities and the dereddened one for which high extinction of $A_{V}=3.3$~mag is assumed. Dereddening increases the flux in both bands, but much more significantly in the F336W band. \emph{Right:} accretion luminosity and accretion rate results as a function of the assumed extinction magnitude. The orange dashed line shows that the trend for accretion luminosity scales as $10^{0.4A_{\halpha}}$, which was assumed in \citet{Wagner2018}, \citet{Aoyama2019}, and \citet{Hashimoto2020}. Because reddening is stronger at shorter wavelengths, accretion luminosity estimates increase with $A_{V}$ at a much greater rate in our analysis (blue solid line) that calculates accretion excess emission in both UV and \halpha. In the high extinction scenario \citep{Hashimoto2020}, extinction increases the $\Lhalpha+\Lcont$ estimate for \Lacc (blue line) by nearly one order of magnitude stronger compared to the scenario in which only $A_{\halpha}$ is considered.}
    \label{fig:extinction}
  \end{figure*}

 The lack of a tight constraint on the line-of-sight extinction of \pdsb leads to the most significant systematic uncertainty in our \mdot estimate. Because {extinction} is much more effective in the F336W band than at H$\alpha$ ($A_{\mathrm{F336W}}/A_{\mathrm{H\alpha}}=2.0$, assuming $R_{V}=3.1$, \citealt{Cardelli1989}), accretion luminosity estimate increases with extinction at a much steeper rate in our analysis than those solely based on  H$\alpha$ \citep[Figure~\ref{fig:extinction}]{Wagner2018,Aoyama2019,Hashimoto2020}. \edit1{However, working at longer wavelengths does not mean that estimates based on \halpha (or Br$\gamma$ or other lines) are more robust to extinction than the Balmer continuum measurements.  Until an accurate correlation between \halpha and accretion continuum luminosity can be confidently measured for planets, the uncertainty in extinction similarly affects all luminosity estimates. } In the high extinction scenario ($A_{V}>3$~mag) adopted in \citet{Hashimoto2020},  the mass accretion rate of PDS70 b will be $\dot{M}{\sim}7\times10^{-7}M_{\mathrm{Jup}}\,\mathrm{yr}^{-1}$, two orders of magnitude greater than the result assuming zero extinction.


  \edit1{Previous observations found that neither interstellar nor circumstellar material is likely to cause extinction to \pdsb \citep{Mesa2019a, Wang2020}. By fitting PDS 70's SED, \citet{Wang2020} placed a $3\sigma$ upper limit of $A_{V}=0.15$~mag for the line-of-sight extinction of the star. On the other hand, circumplanetary material can introduce significant extinction, but the evidence is inconclusive for this system. \citet{Hashimoto2020} argued for an extinction of at least $A_{\mathrm{H\alpha}}{\sim}2.4$ mag ($A_{V}{\sim}3.3$ mag) for \pdsb, based on a nondetection of H$\beta$ emission from \pdsb and a theoretical H$\beta$-H$\alpha$ flux ratio. However, applying a similar method to the nondetection of Pa$\beta$ emission, \citet{Uyama2021} found a low extinction for \pdsb, otherwise its Pa$\beta$ emission should have been detected.  \citet{Wang2021} showed that the best-fitting extinction value to the 1 to 5 \micron{} SED of \pdsb is dependent upon the model grid. Because extinction affects the UV observations much more significantly than optical and IR, our F336W detection of \pdsb can potentially offer a tight constraint when an accurate theoretical UV flux prediction of \pdsb is available.
  }

  \subsection{Accretion-induced Emission: Line versus Continuum}
  \begin{figure}
    \centering
    \includegraphics[width=\columnwidth]{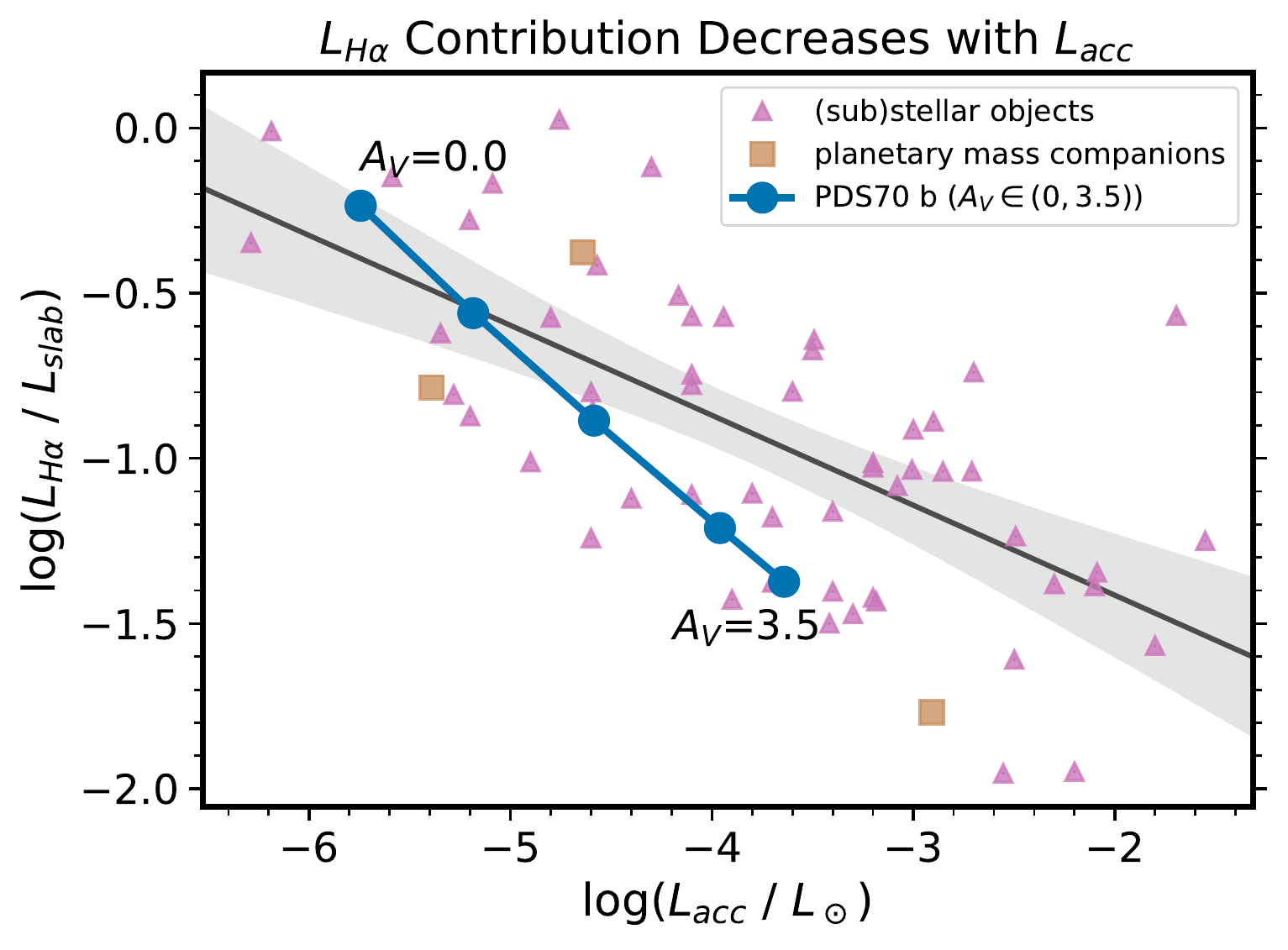}
    \caption{The relation between \Lhalpha and \Lacc for PDS70b (this work), planetary-mass companions \citep{Zhou2014}, brown dwarfs, and low-mass ($<0.2\,M_{\odot}$) stars \citep{Herczeg2008, Herczeg2009,Rigliaco2012,Alcala2014, Alcala2017}. The blue line tracks \Lhalpha/\Lacc for PDS70b for $A_{V}$ from 0 to 3.5. The gray line is the linear regression of $\log(\Lhalpha/\Lacc)$ and $\log(\Lacc)$, indicating a trend that low accretion luminosity objects release a greater proportion of their accretion energy in the \halpha line. Our measurements for PDS70b agree well with the general trend. In particular, the $A_{V}=0$ value falls within the expected range based on this linear regression.}
    \label{fig:halpha_cont}
  \end{figure}

  \pdsb{}'s \halpha{}-to-continuum accretion luminosity ratio is $\Lhalpha/\Lcont=0.56$. Such a high \halpha contribution to the total accretion excess emission is similar to those observed in the accreting planetary-mass companions GSC06214-0210b and DH Tau b \citep{Zhou2014}, as well as a few other slowly accreting substellar objects \citep[Figure \ref{fig:halpha_cont},][]{Herczeg2008, Herczeg2009, Rigliaco2012, Alcala2014, Alcala2017}. This ratio decreases if a greater extinction magnitude is assumed. Dereddening the photometry with an extinction law of $A_{V}=2\,\mathrm{mag}$ ($R_{V}=3.1$) reduces $\Lhalpha/\Lcont$ to ${\sim}10\%$.

\pdsb{}'s high \halpha emission contribution to the total accretion luminosity is likely rooted in its planetary mass and relatively weak accretion flow \citep{Aoyama2018}. Compared to stars,  \pdsb{}'s accretion shock likely has a lower hydrogen number density \citep{Aoyama2019, Hashimoto2020}, resulting in optically thin emission with stronger line emission and high Balmer jump \citep[e.g.,][]{Herczeg2009}. In addition, \citet{Aoyama2019} found that in planetary accretion shocks,  \halpha{} is emitted from the post-shock gas. This is in contrast to the stellar magnetospheric accretion scenario where \halpha{} comes from  the pre-shock flows \citep[e.g.,][]{Muzerolle1998,Alcencar2012}. This difference further increases the line-to-continuum luminosity ratio in planetary accretion emission.  Our observations directly confirm the divergence between the accretion of \pdsb and accretion in young stellar objects.

  Our results also reinforce the danger of extrapolating empirical stellar $\Lhalpha-\Lacc$ relations \citep{Rigliaco2012, Ingleby2013, Alcala2014} into the planetary accretion regimes, which has been previously identified in theoretical studies \citep[e.g.,][]{Aoyama2018,Thanathibodee2019,Szulagyi2020}. Those linear relations cannot fully capture the diversity in the mechanisms of accretion excess emission between the planetary and stellar scenarios. Our measurements of both $\Lhalpha$ and $\Lcont$ offer a more direct way to probe the mass accretion rate of PDS70b. The high $\Lhalpha/\Lcont$ ratio of \pdsb significantly improves its star-to-planet contrast in the \halpha band, which may serve as further motivation to search for accreting giant planets with \halpha \citep[e.g.,][]{Zurlo2020,Close2020}
  

  \subsection{A Comparison to Measurements Based on Accretion Models}
  \edit1{We compare our accretion rate result, which is a direct translation of the observed UV and \halpha fluxes, to those derived by modeling \halpha emission from \pdsb's accretion shocks \citep{Aoyama2018, Thanathibodee2019, Hashimoto2020, Aoyama2020}. These models predict theoretical correlations between the line emission and total accretion luminosities so the accretion rate measurements do not depend upon empirical correlations established for stars. In these studies, results vary based on the model assumptions. \citet{Thanathibodee2019} expanded the magnetospheric accretion model \citep{Muzerolle1998}, in which line and continuum are from different regions,  to planetary-mass regimes. They found an accretion rate of $1\,\mbox{to}\, 1.5\times10^{-8}\,M_{\mathrm{Jup}}\,\mathrm{yr^{-1}}$ for PDS 70 b. By contrast, \citet{Aoyama2018} and \citet{Aoyama2020} modeled the planetary accretion shock and found that the post-shock gas is the primary emitter of both the line and continuum flux. Most recently, they reported $1.1\,\mbox{to}\,8.0\times10^{-8}\,M_{\mathrm{Jup}}\,\mathrm{yr^{-1}}$ (the range reflects the \halpha flux difference in ground-based observations, see\S\ref{sec:time_resolved}).}

  \edit1{Because these model-based estimates depend on the adopted \Lhalpha and there is a degeneracy between \mdot and the adopted planetary mass and radius (Equation~\ref{eq:mdot}), we scale those accretion rates so that they are all based on the same values as our measurements: $\Lhalpha=6.5\times10^{-7}\,L_{\mdot}$, $M_{\mathrm{p}}=1M_{\mathrm{Jup}}$ and $ R_{\mathrm{p}}=1.75R_{\mathrm{Jup}}$.  This leads to $\mdot=8\times10^{-8}M_{\mathrm{Jup}}\mathrm{yr^{-1}}$  and $\mdot=5\times10^{-7}M_{\mathrm{Jup}}\mathrm{yr^{-1}}$ for the \citet{Thanathibodee2019} and \citet{Aoyama2020} models, respectively. They are both significantly higher than our measurement of $\mdot=1.4\pm0.2\times10^{-8}M_{\mathrm{Jup}}\mathrm{yr^{-1}}$.}

  \edit1{This difference could suggest that the production of H$\alpha$ emission from the accretion shock  is more efficient than what the models predict, i.e.,  producing the observed H$\alpha$ luminosity requires a lower accretion rate than what these models require. Alternatively, our accretion rate may be underestimated, as we neglect emission lines other than \halpha{}. If \pdsb's Ly$\alpha$ line emission is indeed more than one order of magnitude stronger than H$\alpha$, as suggested in \citet{Aoyama2018, Aoyama2020}, its unaccounted emission can explain our low accretion rate.
  }

  \subsection{The Mass Accretion Rate of PDS 70 b in Context}

  The mass accretion rate of young stellar objects  follows a power law in stellar mass, albeit with considerable uncertainty. With our mass accretion measurement, we can test whether \pdsb, a planet in the gap of a protoplanetary disk, follows the same $\mdot\mbox{ vs. }M$ trend as stars. Extending the power law in \citet{Hartmann2016} to the mass of PDS70 b yields an accretion rate of $6\times10^{-11}M_{\mathrm{Jup}}\,\mathrm{yr^{-11}}$, two orders of magnitude lower than our measurement. Considering a 0.75 dex intrinsic scatter and a 0.5 dex uncertainty due to age (Equation 12 in \citealt{Hartmann2016}), we still find \pdsb's accretion rate significantly higher than the extrapolated stellar relation. \pdsb's high accretion rate can be explained by the fact that \pdsb is embedded in a mass reservoir that is constantly mass-loading its disk, and the stars in comparison have detached from their envelopes. This hypothesis can be tested by a future comparison between \pdsb's accretion rate to the $\mdot\mbox{-}M$ trend for Class I young stellar objects that are still embedded in their envelopes. \edit1{Alternately, if there is a population of accreting gas giants that have highly variable accretion rates, we would detect the bright outliers and it would not be surprising that the one we are analyzing is much brighter than the rest.}

  Based on our measurement,  \pdsb{}'s accretion rate is less than 20\% of the lowest value found for \pds \citep[$0.6\,\mathrm{to}\,2.2\times10^{-7}M_{\mathrm{Jup}}\mathrm{yr^{-1}}$][]{Thanathibodee2020}. Therefore, accretion onto the star dominates the mass flow within the circumstellar disk of the PDS70 system. Compared to the average mass accretion rate of $3\mbox{ to }8\times10^{-7}M_{\mathrm{Jup}}\mathrm{yr^{-1}}$ \citep{Wang2020} over its formation period of ${\sim}$5~Myr, the current mass accretion of \pdsb is about two orders of magnitude lower. \pdsb is likely in a relatively quiescent accretion state and {may have} gained most of its mass during accretion outburst periods \citep[also see ][]{Brittain2020} . However, these interpretations rely on our assumptions for \Lacc and mass accretion rate estimates. For example, under a high extinction assumption, the observed \mdot of \pdsb will be approximately the same as its average accretion rate, suggesting the planet is rapidly growing while the star is only weakly accreting. We expect such divergence of interpretations to be minimized by a joint effort of multiwavelength observations of \pdsb and modeling of line and continuum emission in planetary accretion shocks \citep{Aoyama2020}.


  \section{Conclusions}

  By applying a suite of image reconstruction and angular differential imaging strategies \citep{Lauer1999,Rajan2015} to HST/WFC3/UVIS observations, we have successfully detected the young giant exoplanet \pdsb in both the F336W (UV) and F656N (H$\alpha$)  bands. This is the first direct detection of an exoplanet in the UV and offers the first constraint on the accretion excess emission at the Balmer jump for an exoplanet. Our findings are as follows:

  1. The signal-to-noise ratios of \pdsb's detections in the F336W and F656N bands are 5.3 and 7.8, respectively. The positions of the two detections agree with each other within $1\sigma$/15mas in both radial and tangential directions.

  2. Neither band yields a $>3\sigma$ detection for \pds~c. At its expected position in the F656N image, we find a $1.4\sigma$ signal, corresponding to a \halpha flux of $2.6\pm 1.9 \times 10^{-16} \mathrm{erg}\,\mathrm{s}^{-1}\mathrm{cm}^{-2}$, consistent with the literature values.

  3. The flux densities of PDS70b in the F336W and the F656N bands are $1.4\pm0.3\times10^{-19} \flamunit$ and $9.2\pm1.3\times10^{-17}\flamunit$, respectively. They correspond to hydrogen continuum and \halpha luminosities of $\Lcont=1.2\pm0.2\times10^{-6}\Lsol$ and $\Lhalpha=6.5\pm0.9\times10^{-7}\Lsol$, and a total accretion luminosity of $\Lacc=1.8\pm0.2\times10^{-6}\Lsol$. Under a no-extinction assumption, the \halpha luminosity accounts for up to 56\% of the continuum luminosity and approximately 36\% of the total accretion luminosity. The high contribution of \halpha line emission to the total accretion luminosity reinforces the trend that the accretion shocks in accreting planetary-mass objects produce a greater portion of \halpha line emission than in stars.

  4. The accretion luminosity corresponds to a mass accretion rate of $\mdot=1.4\pm0.2\times10^{-8}M_{\mathrm{Jup}}\mathrm{yr^{-1}}$. This result is low compared to previous estimates based on accretion shock modeling of the \halpha line emission \citep{Aoyama2018,Thanathibodee2019,Hashimoto2020}. The discrepancy suggests that either \halpha production in planetary accretion shocks is more efficient than these models predicted, or we underestimated the accretion luminosity/rate. \edit1{By combining our observations with planetary accretion shock models that predict both UV and \halpha flux, we can improve the accretion rate measurement and advance our understanding of the accretion mechanisms of gas giant planets.}

5. Our \halpha flux falls between two  previous ground-based measurements \citep{Wagner2018,Haffert2019,Hashimoto2020} and is over $3\sigma$ higher than the latest and the more precise one \citep{Hashimoto2020}. Our low-cadence (${\sim}$1 month), $\sim30\%$ precision \halpha light curve does not show evidence for variability.

6. Our observations demonstrate that HST/WFC3/UVIS ADI observations can reach a contrast of $3\times10^{-4}$ at 170mas (${\sim}6\lambda/D$) in the UV. HST is complementary to ground-based extreme adaptive optics facilities in direct-imaging exoplanets. In particular, HST observations will be most effective in detecting planets around faint host stars, probing variability in planetary accretion, and constraining the UV continuum emission from accreting giant exoplanets.

\acknowledgments We thank the referee for a prompt report. The authors thank Dr. Gabriel-Dominique Marleau and Dr. Yuhiko Aoyama for enlightening discussions. Y.Z. thanks Dr. Feng Long and Dr. Rixin Li for their support and encouragement during the lockdown. B.P.B. acknowledges support from the National Science Foundation grant AST-1909209. L.C. is partially supported by NASA XRP grant  80NSSC18K0441. G.J.H. is supported by general grant 11773002 awarded by the National Science Foundation of China. Support for Program number 15830 was provided by NASA through a grant from the Space Telescope Science Institute, which is operated by the Association of Universities for Research in Astronomy, Incorporated, under NASA contract NAS5-26555. Based on observations made with the NASA/ESA Hubble Space Telescope, obtained in GO program 15830 at the Space Telescope Science Institute.

\appendix
\counterwithin{figure}{section}
\section{Robustness Tests for the UV Detection of PDS 70 b}
We conduct additional robustness tests for the F336W/UV band detection. Our tests result in five independent lines of evidence  supporting the robustness of the detection.
  
  1. The detection is not driven by a single epoch. Time-dependent systematics or PSF anisotropy may introduce false positive signals. Because our observations consist of six visit sets spanning five months in time and $197^{\circ}$ in total telescope roll angle, such systematics are unlikely to introduce false detections in multiple epochs. If the detection was in fact  due to a strong false positive in one epoch, excluding data from this epoch should eliminate the signal in the final processed frame.

  \begin{figure*}[t]
    \centering
    \includegraphics[width=\textwidth]{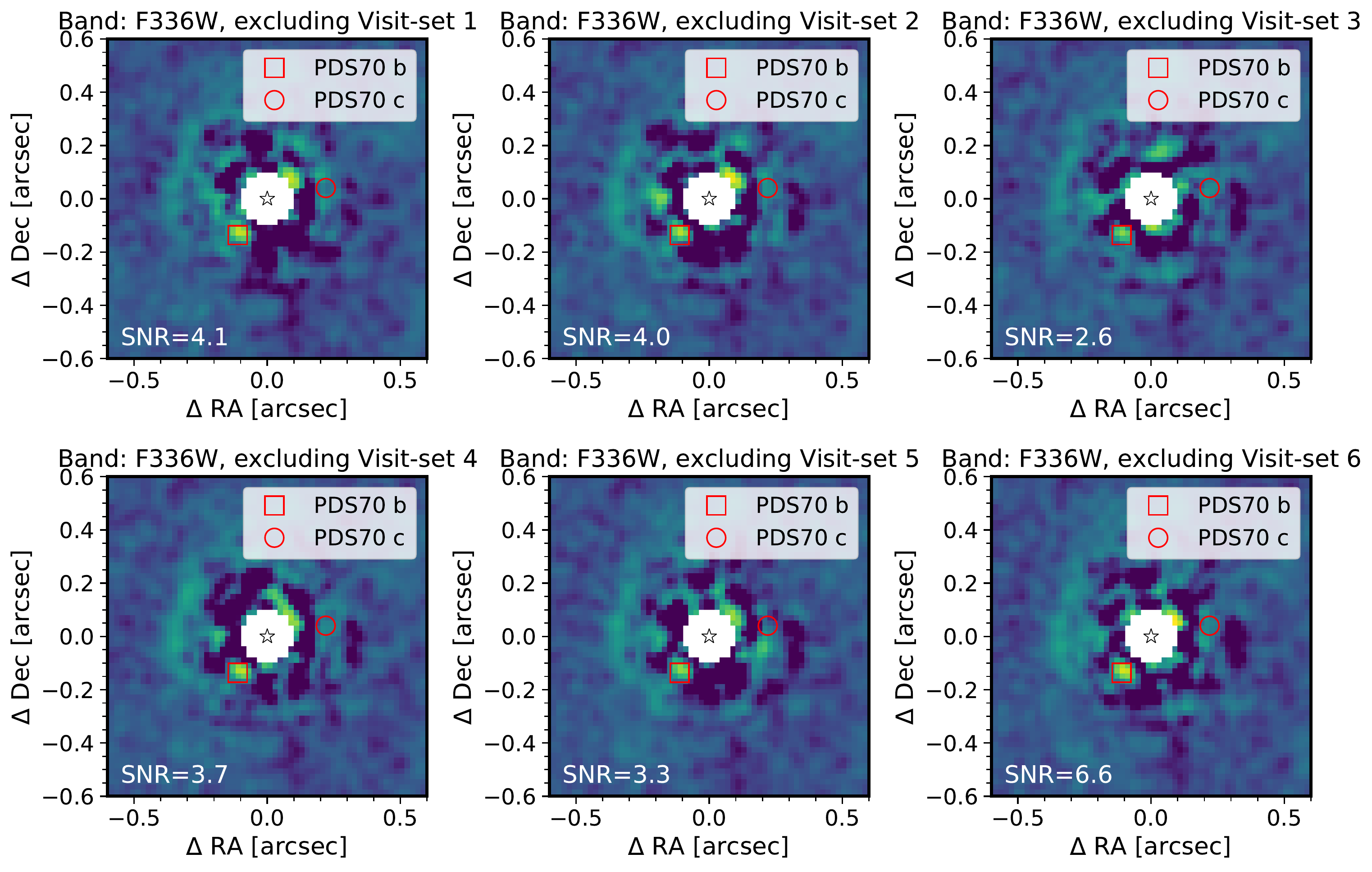}
    \caption{Primary-subtracted images in F336W. Each image is the combination of five out of six visit-sets of observations.  The expected positions of PDS 70 b and c are marked by squares and circles. The detection SNR for PDS 70 b is annotated in each panel. PDS 70 b is detected with SNR$>2.5$ in every case.}\label{fig:leave-one-out}
  \end{figure*}
  
  To test if the detection is driven by a single (outlier) epoch, we make primary-subtracted images with each visit-set in turn excluded from the final combined image. As shown in Figure~\ref{fig:leave-one-out}, PDS70b is detected in every image with SNR$>2.5$, regardless of which visit-set is excluded.  This test result makes it unlikely that the F336W detection is a false positive signal driven by time-dependent systematics or PSF anisotropy.

  2. The detection does not rely on the optimization area of the KLIP algorithms. We experiment with various geometries and optimization areas for KLIP. They include circular apertures centered on the planet with various radii and annular sectors with a variety of inner/outer radii and PA spans. The point source at the expected location of PDS70b is consistently detected regardless of the geometry of the optimization region.

  3. Injected synthetic PSFs with the same flux densities and separations as \pdsb are recovered with similar SNRs to the observed values for \pdsb (Figure~\ref{fig:injection-and-recovery}). 
 
\begin{figure*}
    \centering
    \includegraphics[width=0.48\textwidth]{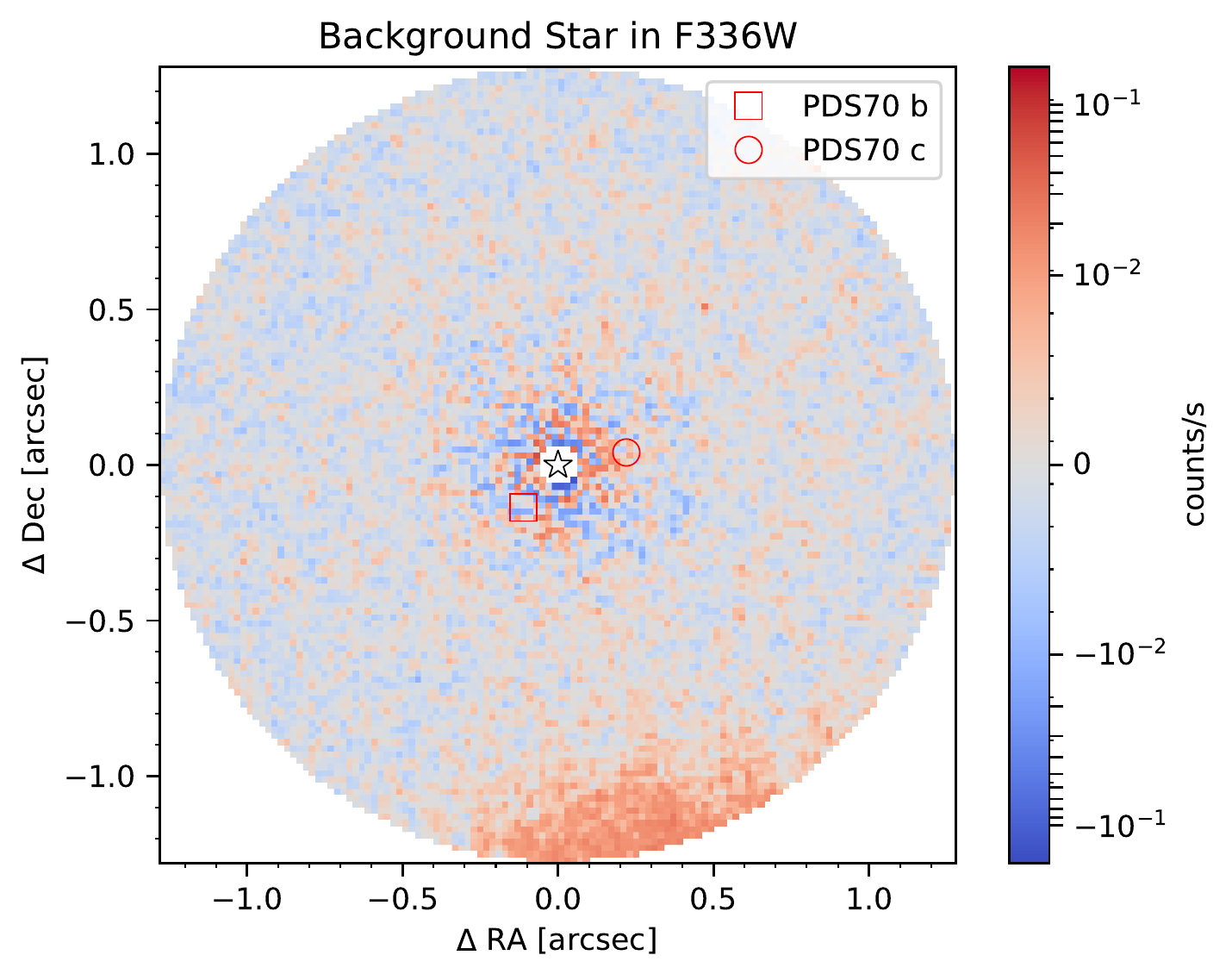}
    \includegraphics[width=0.48\textwidth]{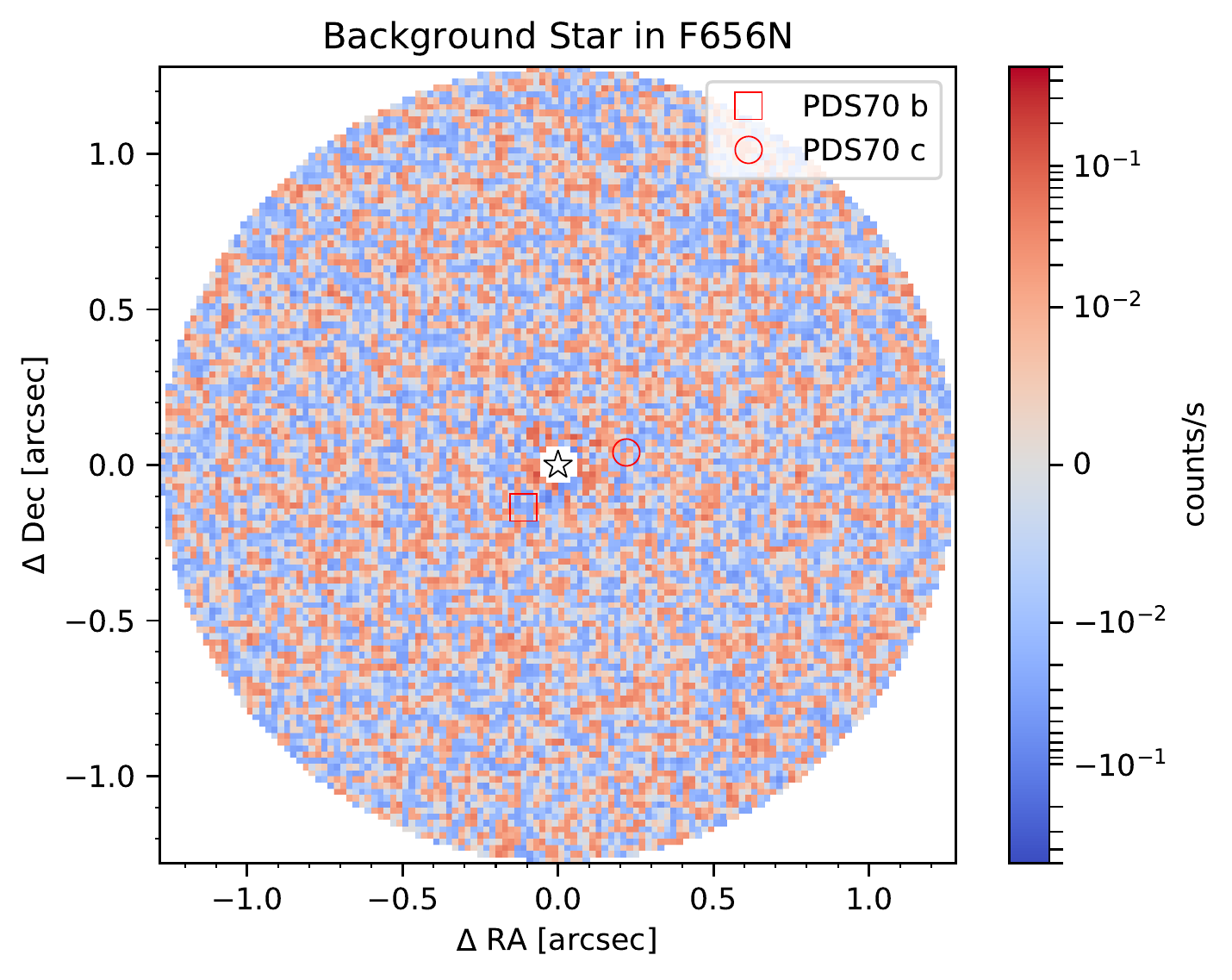}
    \caption{Primary-subtracted images of a close-by ($2.3\arcsec$) background star. No point-like sources are detected in these images.}
    \label{fig:bck}
  \end{figure*}

  4. Similar signals are not present in a PSF-subtracted image of a background star. We apply the data reduction pipeline to a nearby background star in the field of view of our observations.  The KLIP parameters and setups are identical to those in subtracting PDS 70 PSFs. The PSF-subtracted images for the background star do not show any point source like signals (Figure \ref{fig:bck}).

  5. The location of the detected point source in the F336W band is consistent with the one in the F656N band within $1\sigma$ (Figure~\ref{fig:astrometry_comparison}). They also agree with the expected astrometry for PDS70b estimated from previous studies \citep[e.g.,][]{Wang2020}.

  Taken together, these five indicators provide strong evidence that the detected point sources are associated with \pdsb.

\end{document}